# On-surface synthesis of graphene nanoribbons with zigzag edge topology

Pascal Ruffieux[1]\*, Shiyong Wang[1]\*, Bo Yang[3]\*, Carlos Sanchez[1]\*, Jia Liu[1]\*, Thomas Dienel[1], Leopold Talirz[1], Prashant Shinde[1], Carlo A. Pignedoli[1,2], Daniele Passerone[1], Tim Dumslaff[3], Xinliang Feng[4], Klaus Müllen[3†], Roman Fasel[1,5†]

[1] Empa, Swiss Federal Laboratories for Materials Science and Technology, 8600 Dübendorf, Switzerland.

[2] NCCR MARVEL, Empa, Swiss Federal Laboratories for Materials Science and Technology, 8600 Dübendorf, Switzerland.

[3] Max Planck Institute for Polymer Research, 55128 Mainz, Germany.

[4] Department of Chemistry and Food Chemistry, Technische Universität Dresden, 01062 Dresden, Germany.

[5] Department of Chemistry and Biochemistry, University of Bern, 3012 Bern, Switzerland

\*These authors contributed equally to this work

†Corresponding authors. E-mail: roman.fasel@empa.ch (R.F); muellen@mpip-mainz.mpg.de (K.M.)

**Graphene-based nanostructures exhibit a vast range of exciting electronic properties that are absent in extended graphene. For example, quantum confinement in carbon nanotubes and armchair graphene nanoribbons (AGNRs) leads to the opening of substantial electronic band gaps that are directly linked to their structural boundary conditions[1,2]. Even more intriguing are nanostructures with zigzag edges, which are expected to host spin-polarized electronic edge states and can thus serve as key elements for graphene-based spintronics[3]. The most prominent example is zigzag graphene nanoribbons (ZGNRs) for which the edge states are predicted to couple ferromagnetically along the edge and antiferromagnetically between them[4]. So far, a direct observation of the spin-polarized edge states for specifically designed and controlled zigzag edge topologies has not been achieved. This is mainly due to the limited precision of current top-down approaches[5–10], which results in poorly defined edge structures. Bottom-up fabrication approaches, on the other hand, were so far only**

**successfully applied to the growth of AGNRs[11–13] and related structures[14–16]. Here, we describe the successful bottom-up synthesis of ZGNRs, which are fabricated by the surface-assisted colligation and cyclodehydrogenation of specifically designed precursor monomers including carbon groups that yield atomically precise zigzag edges. Using scanning tunnelling spectroscopy we prove the existence of edge-localized states with large energy splittings. We expect that the availability of ZGNRs will finally allow the characterization of their predicted spin-related properties such as spin confinement[17] and filtering[18,19], and ultimately add the spin degree of freedom to graphene-based circuitry.**

To explore the fundamental electronic and magnetic properties related to zigzag edges and to realize specific carbon nanostructures for the controlled manipulation of their spin states, ZGNRs with atomically precise edges are required. For GNRs with armchair edges, it was demonstrated that atomic precision can indeed be achieved by a bottom-up approach based on the surface-assisted polymerization and subsequent cyclodehydrogenation of specifically designed oligophenylene precursor monomers[11]. The on-surface synthesis has been applied by many groups to fabricate a number of different AGNR structures[11–13], N-doped AGNRs[14,15] as well as AGNR heterostructures[15,16]. It is, however, not directly suited for ZGNRs since polymerization of monomers via aryl-aryl coupling does not take place along the zigzag but along the armchair direction (Fig. 1a). In addition, dehydrogenative cyclization of phenyl subgroups is not sufficient to form pure zigzag edges, thus calling for a totally new chemical design. Thereby, additional carbon functions must be placed at the edges of the monomers to complete the tiling toolbox needed for the bottom-up fabrication of arbitrary GNR structures.

Here, we report a bottom-up fabrication approach to ZGNRs. In our unique protocol, surface-assisted polymerization and subsequent cyclization of suitably designed molecular precursors carrying the full structural information of the final ZGNR afford atomic precision with respect to ribbon width and edge morphology. The groundbreaking idea depends upon the choice of a unique U-shaped monomer as **1** shown in Fig. 1b. With two halogen functions for thermally induced aryl-aryl-coupling at the $R_1$ positions, it allows the polymerization toward a snake-like polymer. It is the beauty of this design that additional phenyl groups at the $R_2$ position fill the holes in the interior of the undulating polymer. The crucial precursor is monomer **1a** which carries two additional methyl groups. In such a case, apart from the polymerization and planarization, an oxidative ring closure including the methyl groups is expected which would then establish two new six-membered rings together with the zigzag edge structure. To our delight, this concept could indeed be synthetically realized under reaction monitoring and structure proof by scanning tunneling microscopy (STM) and non-contact atomic force microscopy (nc-AFM), as shown in Fig. 2.

Monomer **1a** was successfully obtained via multi-step organic synthesis (see the Supplementary Information) and deposited on the clean Au (111) single crystal surface by thermal sublimation under ultra-high vacuum (UHV) conditions. If the surface is held at the dehalogenation temperature of 475 K, precursor monomers are immediately activated and undergo polymer formation via radical addition (*step 1*). Further annealing to the cyclodehydrogenation temperature of 625 K is then applied to form the final 6-ZGNRs (*step 2*). This two-steps process has been successfully monitored by STM (Fig. 2a, b). Large-scale STM images of the Au (111) surface after deposition of precursor monomer **1a** at 475 K substrate temperature reveal the formation of long (~ 50 nm) polymers for which the meandering apparent maxima have a periodicity of 1.55 nm, evidencing covalent bond formation between the precursor monomers (Fig. 1c). The maxima with apparent height of 0.3 nm are attributed to the sterically induced out-of-plane conformation of the phenyl ring carrying the methyl groups. Remarkably enough, further annealing the sample to 625 K results in a complete planarization of the linear structures and a decrease in apparent height to 0.2 nm, consistent with the formation of the fully conjugated ribbon structure[11]. Small-scale images (inset of Fig. 2b) reveal completely smooth and flat edge areas. This indicates that besides the cyclodehydrogenation of the two phenyl rings also the methyl groups are dehydrogenatively incorporated to form a fully conjugated system with atomically precise zigzag edges. Further structural details are accessible by nc-AFM imaging with a CO-functionalized tip (Fig. 2c), which allows for a direct imaging of the local bond configurations at small distances[20]. The achieved resolution directly confirms that width and edge morphology correspond to the expected 6-ZGNR structure as defined by the design of **1a**. Furthermore, we can unambiguously state that the zigzag edge atoms have the expected mono-hydrogen termination. Other possible terminations such as radical edges due to complete dehydrogenation or $H_2$ termination can be discarded due to the absence of bending across the ribbon (related to bonding of the radical edges to the substrate[21]) and the absence of the distinct maxima related to $H_2$ edge termination (see Supporting Figure S1), respectively. Thus, our 6-ZGNRs exhibit atomically precise edges with the expected CH termination. Both features are crucial to host the predicted antiferromagnetic edge states.

As can be seen from the STM image in Fig. 2b, the chemistry of the ZGNR fabrication process faces intrinsic complications such as frequent thermally induced chemical cross-linking of ribbons during the cyclization step, and, most severely, a strong electronic coupling between the ribbons and the metal surface which obscures the detection of the electronic edge states. In fact, no evidence for increased intensity at the zigzag edges could be obtained in differential conductance (dI/dV) maps recorded with a tunneling resistance down to 0.6 MΩ, and spectra taken above the 6-ZGNRs are dominated by the (somewhat up-shifted) surface state of the underlying Au(111) substrate (see Supporting Figure S2). First fingerprints of the 6-ZGNR edge states could be obtained in dI/dV maps taken with a special tip (of unknown termination) that featured a high density of states at the Fermi energy (see Supporting Figure S2). Clear and unambiguous evidence for the sought-after edge states can, however, be obtained for 6-ZGNRs manipulated with the STM tip onto post-deposited insulating NaCl islands, where they are electronically decoupled from the underlying metal substrate[22]. Figure 3a shows the example of a STM image of a 6-ZGNR bridging between two NaCl islands. Figure 3d displays a dI/dV spectrum taken at the edge of the decoupled ZGNR segment. In sharp contrast to the result on Au(111), the spectrum clearly exhibits three resonance peaks near the Fermi level, with an energy splitting of $\Delta^0 = 1.5$ eV and $\Delta^1 = 1.9$ eV between the two occupied states and the unoccupied one, respectively (Fig. 3b). dI/dV maps acquired at these peaks document that the corresponding states are highly localized at the zigzag edges (Fig. 3d). Their characteristic features, such as a protrusion at each outermost zigzag carbon atom and an enhanced intensity at the ribbon terminus, are in excellent agreement with the local density of states of the corresponding Kohn-Sham density functional theory (DFT) orbitals given in Fig. 3e. While effective mean-field theories, such as Kohn-Sham DFT, tend to provide reliable information about the energy level ordering and the shape of orbitals in graphene nanostructures, the same is not true for the size of the electronic gap. More accurate predictions of the band gaps of graphene nanoribbons can be obtained using the $G_0W_0$ approximation of many-body perturbation theory[23]. The resulting quasiparticle band structure and the corresponding DOS are presented in Fig. 3c. The size of the energy splittings $\Delta^0 = 1.4$ eV and $\Delta^1 = 1.7$ eV are in good agreement with experiment. While edge states have previously been observed in a number of systems with less well-defined zigzag edges[9,24–27], the reported energy splittings vary greatly and are significantly smaller than in the present case. This indicates that the electronic structure of zigzag edges is extremely sensitive to edge roughness and interaction with the supporting substrate.

The first synthesis of GNRs with perfect zigzag edge periphery and convincing proof of their edge states opens up new vistas not only with respect to the experimental verification of the intriguing electronic, optical and magnetic properties predicted by theory[28–30], but also their systematic engineering via modified ZGNR types. In a first step along these lines, we further refined our monomer design by introducing an analogous compound **1b** (Fig. 4) which is similar to **1a** except that it bears an additional phenyl group at the $R_3$ position (see the Supporting Information). We expected that due to the steric hindrance brought about by the twisted phenyl group, the growing ribbons would be more efficiently decoupled from the surface and potentially better shielded from neighboring GNRs. Figure 4b and 4c report the structural characterization of the ZGNRs obtained from monomer **1b** in an analogous thermally induced polymerization – cyclization procedure. To our surprise, the nc-AFM images clearly reveal that at the cyclodehydrogenation temperature of 573 K the external phenyl group undergoes a ring closure under the formation of a fluoranthene type subunit with an incorporated five membered ring, as illustrated in Fig. 4a. Since the additional ring closure can occur by dehydrogenation of either of the neighboring zigzag edge carbon atoms, no fully periodic arrangement of the fluoranthene subunits can be expected. This is confirmed by the nc-AFM image shown in Fig. 4c, which reveals the possible three, four, and five zigzag cusps that separate neighboring fluoranthene subunits.

Controlled edge modification is an immensely attractive strategy for engineering the band structure of edge states. In the present case, however, our emphasis lies on the reduction of ZGNR-substrate interaction. Indeed, the edge modification discussed above reduces the ZGNR-substrate interaction sufficiently for allowing the STM to map the typical features of the edge state. Figure 4d and 4f show constant-height differential conductance maps acquired at –0.15 eV and 0.15 eV, respectively, which reveal increased intensity along the ZGNR edges. While the edge states at pristine zigzag edges occupy exclusively one carbon sublattice (*e.g.* sublattice A),[4] in the case of the edge-modified 6-ZGNR the five-membered ring of the fluoranthene subunit locally disturbs the bipartite character of the graphene lattice by directly connecting carbon atoms belonging to the same sublattice (see the Supporting Figure S3). This topological defect breaks translational symmetry along the zigzag edge and gives rise to a linear combination of Bloch states with nodes at the defects (Fig. 4d,e) and nodes in between defects (Fig. 4f,g), as can be seen both in STS experiments and DFT-based simulations. The energetic ordering of these "edge bands" localized on unmodified edge sections and "defect bands" localized on the pentagon-decorated section of the edge [31] is determined by a delicate interplay between kinetic and Coulomb energy contributions and will be the subject of future work.

The successful bottom-up synthesis of atomically precise ZGNRs opens tremendous opportunities and challenges for the analysis of physical properties (band structure, magnetism, charge/spin transport, etc.) and for the fabrication of ZGNR based devices such as the proposed spin valves[18]. However, the strong interaction of the pristine 6-ZGNR with the metal substrate (as reflected in the obstruction of the edge states' spectral features) raises important questions regarding the chemical reactivity of the zigzag edges, which needs to be controlled in order to study and apply these materials under ambient conditions. The present work is believed to be a milestone case of surface chemistry, which only becomes possible by a combination of creative chemical design and *in-situ* STM monitoring of surface bound reactions.

**Supplementary Information** is linked to the online version of the paper at www.nature/com/nature.

**Acknowledgments:** This work was supported by the Swiss National Science Foundation, the Office of Naval Research BRC Program, the European Research Council (grant NANOGRAPH), the DFG Priority Program SPP 1459, the Graphene Flagship (No. CNECT-ICT-604391), and the European Union Projects UPGRADE, GENIUS, and MoQuaS. We acknowledge the Swiss Supercomputing Center (CSCS) for computational resources (project s507). We thank A. Ferretti for his contribution to this project and Oliver Gröning for valuable discussions.

**Methods:**

**1. Sample Preparation**

Experiments have been carried out under ultra-high vacuum conditions (base pressure $10^{-11}$ mbar) with a low-temperature STM and a low-temperature STM/AFM, both from Omicron-Oxford. A Au(111) single crystal has been used as substrate for the growth of 6-ZGNR and edge-extended 6-ZGNR. The Au(111) surface has been cleaned by repeated cycles of argon ion bombardment and annealing at 750 K for 15 minutes until a clean surface is obtained, as judged by STM. Molecular precursors have been thermally deposited on the clean Au(111) surface held at room temperature with a typical rate of 1 Å/min. After deposition, the sample has been post-annealed in four steps to 475, 525, 575 and 625 K for typically 15 min per step in order to achieve long, high-quality GNRs.

**2. Imaging, manipulation and spectroscopy**

**A**. Constant current STM imaging

STM images have been acquired in the constant current mode at sample temperatures of 77 K or 5 K, as indicated in each case. Scanning parameters are specified in each figure caption.

**B**. Constant height non-contact AFM imaging

Non-contact AFM measurements have been performed with a tungsten tip attached to a tuning fork sensor[32]. The tip has been a posteriori functionalized by the controlled adsorption of a single CO molecule at the tip apex from the previously CO dosed surface [33]. This procedure allows to image the chemical structure of organic molecules [20]. The sensor has been driven close to its resonance frequency (~23570 Hz) with a constant amplitude of approximately 70 pm. The shift in the resonance frequency of the tuning fork (with the attached CO-functionalized tip) has been recorded in constant height mode (Omicron Matrix electronics and HF2Li PLL by Zurich Instruments).

**C**. Transfer of GNRs onto NaCl monolayer islands

We have developed a routine to transfer bottom-up fabricated GNRs from the metal surface onto insulating NaCl islands in order to be able to access their intrinsic electronic structure. This method, in which physisorbed individual 6-ZGNRs are laterally and/or vertically manipulated on the Au(111) surface, consists of four steps: (1) deposition of NaCl islands (thermal evaporation of submonolayer coverage, deposition temperature ~1000 K) on the 6-ZGNR / Au(111) surface; (2) pick-up of one end of a GNR by approaching and retracting the STM tip with low bias (~ -50 mV); (2) lateral displacement of the tip, together with the GNR, above the NaCl island; (3) release of the ribbon by a 3.0 V voltage pulse, leaving the GNR partially adsorbed on NaCl and partially on the metal surface.

**D**. Differential conductance spectroscopy

The differential conductance dI/dV measurements have been performed in a low-temperature STM at 5 K via lock-in technique, using a bias voltage modulation of 20 mV and a frequency of 860 Hz. dI/dV maps have been acquired in the constant-current mode for the decoupled 6-ZGNR, and in the constant-height mode for ribbons on Au(111).

**3. Theoretical details**

DFT calculations have been carried out using the CP2K code [34] for geometry optimizations, and the q-Espresso code [35] for scanning tunneling spectroscopy (STS) simulations. In CP2K, the core electrons and nuclei are represented using the pseudo-potential recommended by Goedecker, Teter and Hutter (GTH) [36] and the valence electrons are treated with a triple-ζ valence basis set with two sets of p-type or d-type polarization functions (TZV2P) [37], and in q-Espresso a plane-wave basis set and norm-conserving pseudopotentials are used. The exchange-correlation has been treated using PBE functional [38].

The edge-extended 6-ZGNR structures shown in Figure 4 were set up starting from the optimized atomic structure of the 6-ZGNR (lattice parameter 2.461 Å) and contain 24 unit cells. A 19 Å (15 Å)-wide region of vacuum was included along the transverse (perpendicular) directions in order to avoid interactions between periodic replicas. Forces on the nuclei were then relaxed until all forces dropped below 5 meV / Å. Using the optimized atomic structure, a self-consistent field calculation was carried out in q-Espresso, using energy cutoffs of 120 Ry and 480 Ry for the wave function and the charge density, respectively, and a k-point grid of $3 \times 1 \times 1$, including the Γ-point.

Quasi-particle corrections have been computed within the framework of many-body perturbation theory, using the $G_0W_0$ approximation to the self-energy as implemented in the Yambo code[39]. The electronic structure of the 6-ZGNR from DFT has been recalculated using 60 Ry plane-wave cutoff, 64 k-points in the first Brillouin zone and 250 bands, covering the energy range up to 21 eV above the highest occupied band. The dielectric matrix has been calculated in the random phase approximation with 8 Ry cutoff for the plane-wave basis. $\varepsilon^{-1}$ was evaluated at frequencies $\omega = 0$ and $\omega = i\, 2\mathrm{Ry}$ and extended to the real frequency axis using the plasmon-pole model by Godby and Needs [40]. A rectangular Coulomb-cutoff has been employed along the directions perpendicular to the GNR axis as described in Ref. [41].

**Figures**

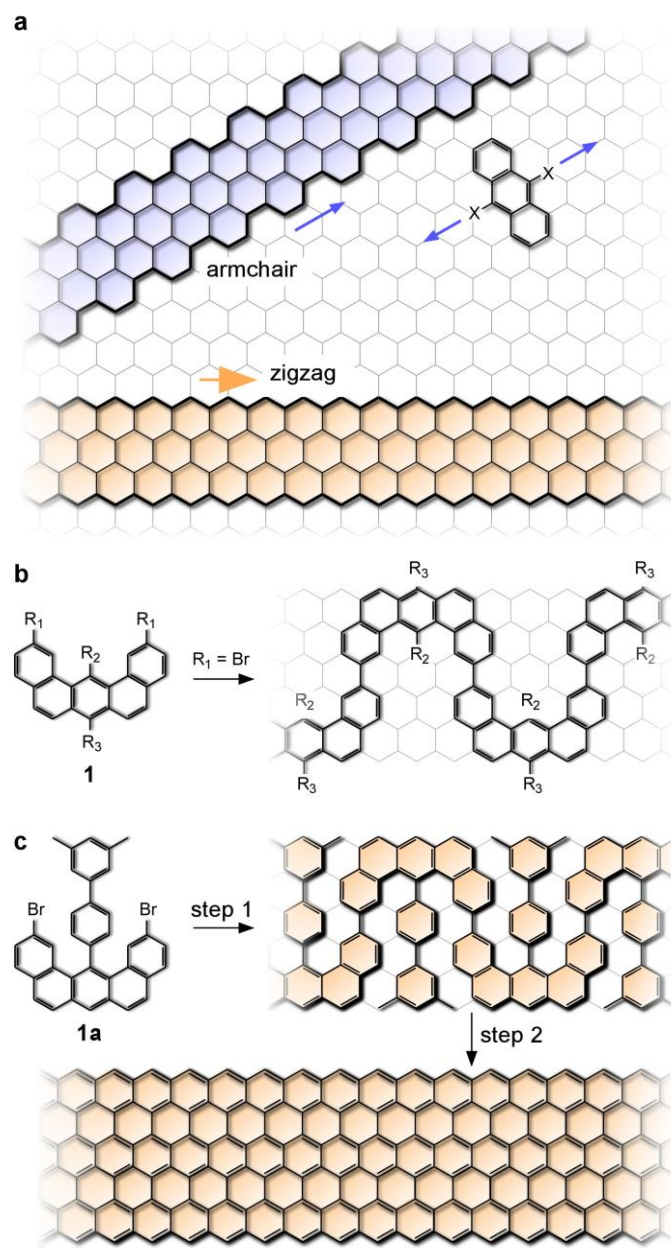

Figure 1

**Fig. 1 | Synthetic strategy to GNRs with zigzag edges. a**, Structure of armchair and zigzag graphene nanoribbons, and exemplary anthracene-based molecular precursor for the bottom-up fabrication of armchair GNRs via aryl-aryl coupling (X = halogen). **b**, U-shaped dibenzo[a,j]anthracene monomer **1** with halogen functions $R_1$=Br designed to allow for surface-assisted aryl-aryl coupling into a snake-like polymer along the zigzag direction. **c**, Monomer **1a**, with an additional dimethyl-biphenyl group in the interior of the U-shape ($R_2$ position), which is designed to afford a 6-ZGNR upon polymerization (step 1) and subsequent cyclization (step 2).

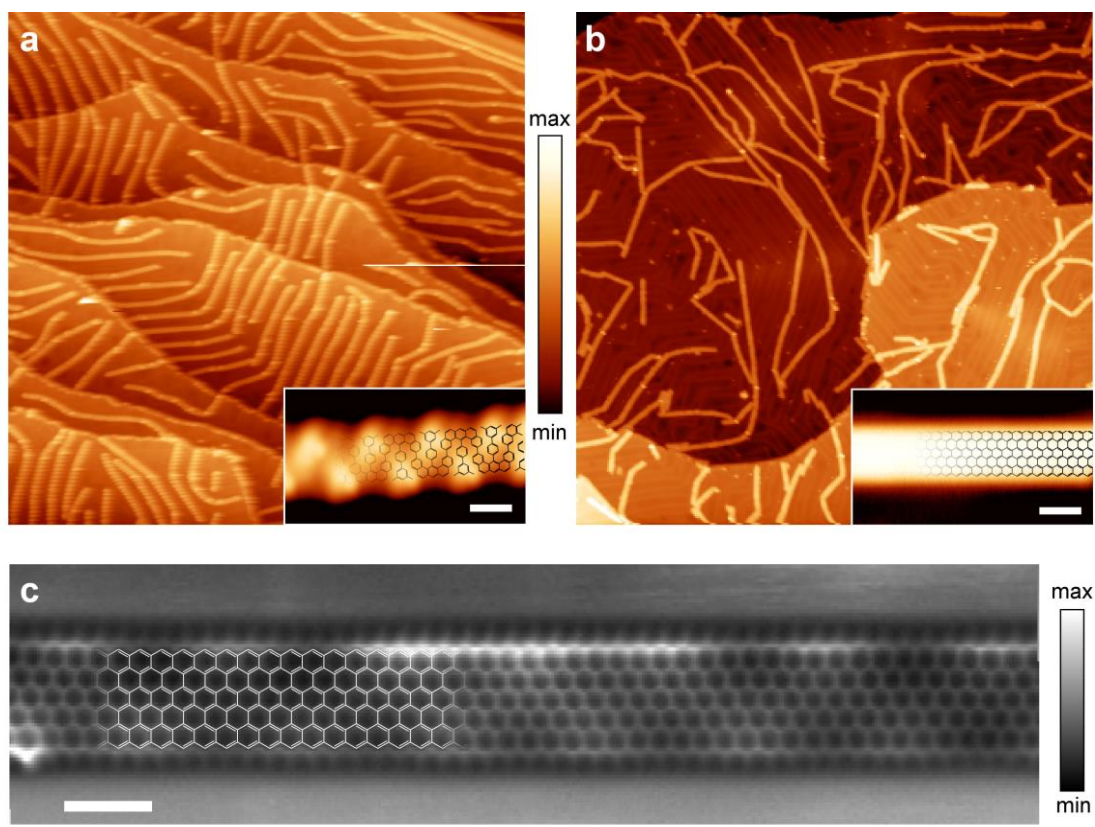

Figure 2

**Fig. 2 | Synthesis and characterization of atomically precise 6-ZGNRs. a**, Large-scale STM image (200 nm x 200 nm) of the Au(111) surface after deposition of monomer **1a** on the surface held at 475 K. Formation of snake-like polymers is observed (V= -1.5 V, I = 40 pA). Inset: High-resolution STM image of the polymer. Zigzag alternation of bright maxima indicates the lifting and/or tilting of the phenyl rings carrying the methyl groups. A structural model is superimposed for comparison. (10.8 nm x 4.2 nm, I = 10 pA, V = -1.3 V). **b**, Large-scale STM image (200 nm x 200 nm) of the Au (111) surface after annealing at 625 K. Flatter appearance, reduced apparent height and no internal structure indicate the complete cyclodehydrogenation of the polymers and the formation of 6-ZGNRs (I = 20 pA, V = -1.0 V). Inset: High-resolution STM of the 6-ZGNRs, which is in excellent agreement with the superimposed structural model (3.7 nm x 12.4 nm, I = 5 pA, V = -0.3 V). **c**, nc-AFM image taken with a CO-functionalized tip. Intra-ribbon resolution shows the formation of 6-ZGNRs with atomically precise CH edges. A $CH_2$ defect is seen in the lower left corner. (4.2 nm x 12.0 nm, $A_{osc}$ = 0.7 Å, sample voltage V = 5 mV).

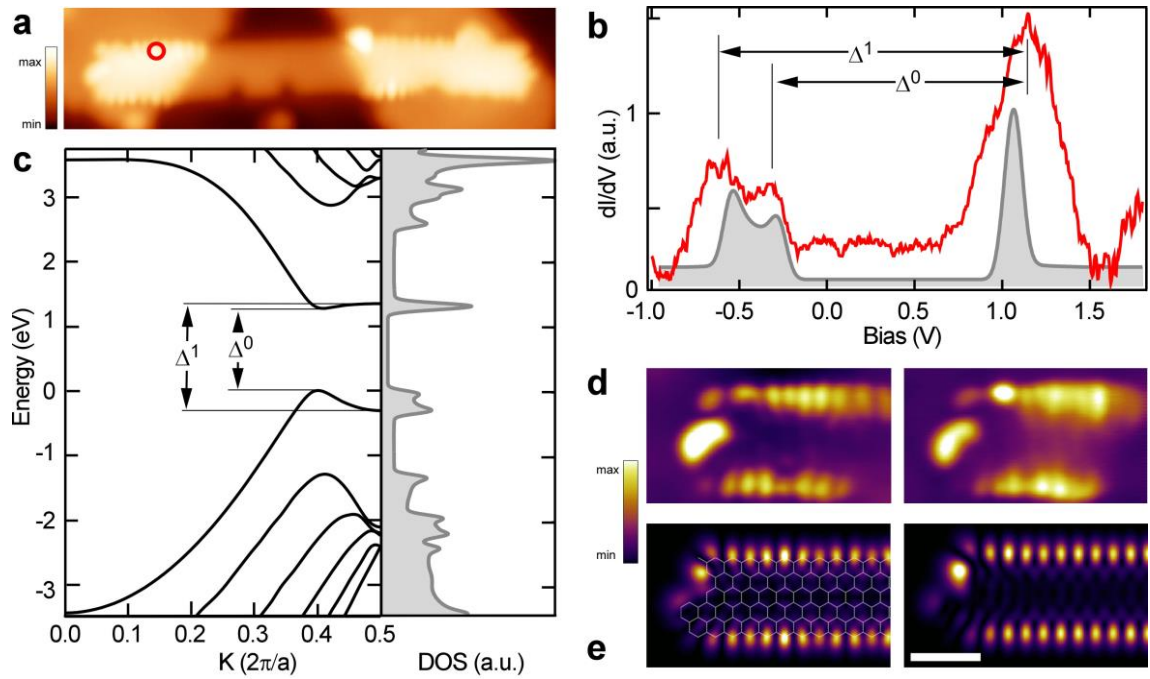

Figure 3

**Fig. 3 | Edge state characterization of 6-ZGNR. a**, STM topography image (12.0 nm x 3.0 nm, U = -0.25 V, I = 100 pA) of a 6-ZGNR bridging between two NaCl monolayer islands, achieved through STM manipulation. **b**, Differential conductance (dI/dV) spectrum taken at the zigzag edge marked by the red dot in **a**. **c**, Quasiparticle band structure (left) and density of states (gray, right panel). **d**, Differential conductance maps of filled (left) and empty (right) edge states taken at a sample bias of -0.3 V and 1.0 V, respectively. **e**, DFT-based local density of states at 4 Å tip-sample distance, showing the spatial distribution of filled (left, with overlaid structural model) and empty (right) edge states (scale bar for **d** and **e**: 1 nm).

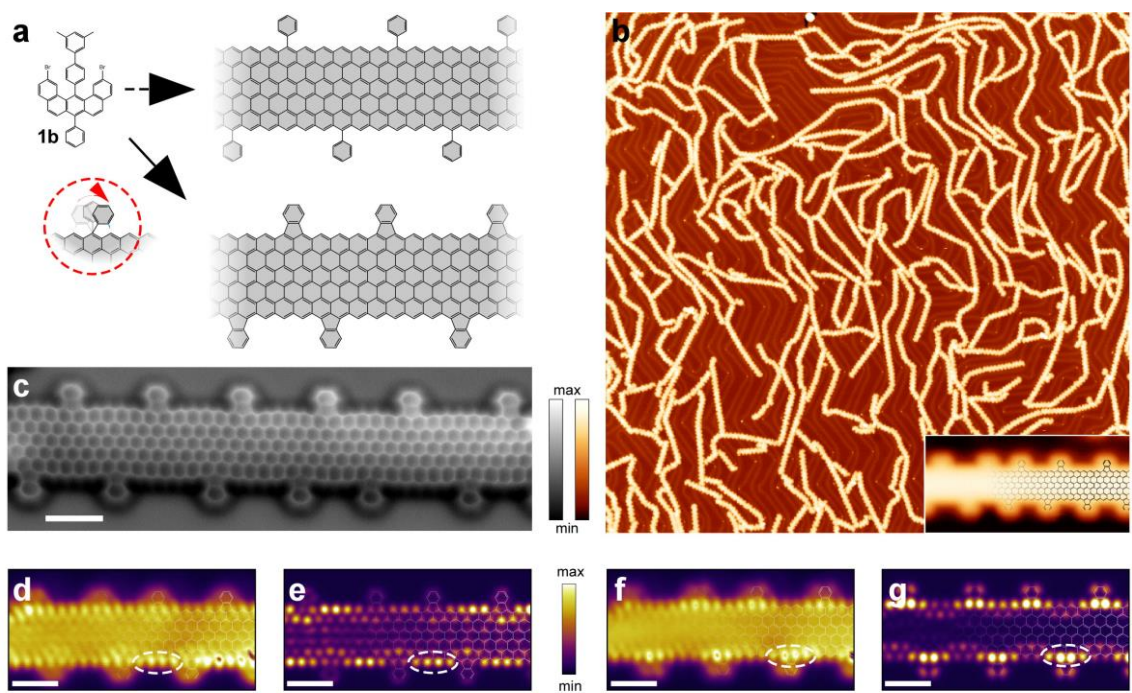

Figure 4

**Fig. 4 | Synthesis and characterization of edge-modified 6-ZGNRs. a**, Monomer **1b**, with an additional phenyl group at the $R_3$ position of the monomer **1a**, which is designed to afford an edge-modified 6-ZGNR upon polymerization and subsequent cyclization. (right) Possible cyclodehydrogenation products assuming no activation of the external phenyl groups (top) and formation of fluoranthene subunits based on an additional dehydrogenative ring closure at the external phenyl groups (bottom). **b**, Overview STM image of edge-modified 6-ZGNR fabricated on a Au(111) surface (200 nm x 200 nm, U = -1.5V, I = 150 pA). The inset shows a high-resolution STM image (7.5 nm x 3.5 nm, U = 0.15V, I = 2 pA). **c**, Constant-height AFM frequency-shift image of edge-modified ZGNR ($A_{osc}$ = 0.7 Å, U = 25 mV). **d**, Constant height dI/dV map of the occupied states (U = -0.15 V). **e**, DFT-based density of states of the highest occupied state. **f**, Constant height dI/dV map of the unoccupied states (U = 0.15 V). **g**, DFT-based density of states of lowest unoccupied level. Dashed ovals in **d** − **g** highlight the zigzag segments with the most prominent contributions for occupied and unoccupied states, respectively. All scale bars: 1 nm.

# Supplementary Information for

## On-surface synthesis of graphene nanoribbons with zigzag edge topology

Pascal Ruffieux, Shiyong Wang, Bo Yang, Carlos Sanchez, Jia Liu, Thomas Dienel, Leopold Talirz, Prashant Shinde, Carlo Pignedoli, Daniele Passerone, Tim Dumslaff, Xinliang Feng, Klaus Müllen\*, Roman Fasel\*

\*Corresponding authors. E-mail: roman.fasel@empa.ch (R.F.); muellen@mpip-mainz.mpg.de (K.M.)

This PDF file includes:

**Synthesis of monomers**

**Figs. S1 to S4**
    S1. $H_2$ defects of 6-ZGNR
    S2. dI/dV spectra and maps of 6-ZGNR taken with a metallic and a 'special' tip
    S3. Geometry frustration at the zigzag backbone
    S4. $^1$H and $^{13}$C NMR spectra

**Supporting references**

**Synthesis of monomers**

A. General information
B. Synthetic route to monomer **1a**
C. Synthetic route to monomer **1b**

A. General information

All reactions dealing with air- or moisture-sensitive compounds have been carried out in the dry reaction vessel under argon protection. Preparative column chromatography has been performed on silica gel from Merck with a grain size of 0.063-0.200 mm (silica gel). Melting points have been determined on a Büchi hot stage apparatus and were uncorrected. NMR spectra have been measured on Bruker DPX 250, AMX 300, and DRX500 spectrometers, and referenced to residual signals of the deuterated solvent. Abbreviations: s = singlet, d = doublet, dd = double doublet, t = triplet, m = multiplet. The high-resolution electrospray ionization mass spectrometry (HR-ESI-MS) has been performed on an ESI-Q-TOF system (maXis, BrukerDaltonics, Germany), where the instrument is operated in wide pass quadrupole mode for MS experiments, with the TOF data being collected between m/z 100–5000. The high-resolution time-of-flight mass spectrometry (APPI-TOF and MALDI-TOF) measurements have been performed on a SYNAPT G2 Si high resolution time-of-flight mass spectrometer (Waters Corp., Manchester, UK) with matrix-assisted laser desorption/ionization (MALDI) or atmospheric pressure photoionization (APPI) source. For MALDI-TOF MS measurement, the samples have been mixed with DCTB ({(2E)-2-methyl-3-[4-(2-methyl-2-propanyl)phenyl]-2-propen-1-ylidene}malononitrile) and dropped on a MALDI sample plate. For APPI TOF MS measurement, the samples have been diluted in toluene to 5 ppm and then infused into the ionization source directly by a Legato 185 syringe pump (KD Scientific, MA, USA) at a flow rate of 5 μL/min. The mass spectrometer has been calibrated against red phosphors under MALDI mode previously and the spectra have been recorded using C60 as lockmass.

Unless otherwise noted, materials have been purchased from Fluka, Aldrich, Acros, ABCR, Merck and other commercial suppliers and used as received without further purification.

B. Synthetic route to monomer **1a**

"U-shaped" monomer **1a** has been synthesized following the synthetic route described in supplementary Scheme 1. As shown in the Scheme 1, compound **2** has been obtained after iodination of 1-bromo-3-methoxybenzene, which had been reported previously[1]. Then, after Sonogashira coupling reaction with (triisopropylsilyl)acetylene, compound **3** has been obtained in 93% yield. Afterwards, compound **4** has been synthesized after borylation reaction of **3** with bis(pinacolato)diboron in 85% yield[2]. In parallel, compound **5** has been prepared by Suzuki coupling of 1,3-dimethyl-5-bromobenzene with 4-(trimethylsilyl) phenylboronic acid in 95% yield. Then, compound **5** has been treated with boron tribromide (BBr$_3$) under neat condition to afford compound **6** in 98% yield[3]. Meanwhile, compound **7** has been synthesized *via* an iodination reaction[4] of **1**, 3-dibromobenzene with iodine in 91% yield. Then, compound **8** has been obtained after a selective Suzuki coupling of **7** with **6** in 48% yield [5].

In the following steps, compound **9** has been prepared after a two-fold Suzuki coupling of **4** with **8** in 48% yield and subsequently treated with tetrabutylammonium fluoride (TBAF) solution for the de-protection reaction to remove triisopropyl (Tips) group [6] to afford compound **10** in 99% yield. The crud compound **10** has been used directly for the next step of reaction. Compound **11** has then been prepared after a catalytic cyclization reaction by treating **10** with platinum (II) chloride catalyst in 63% yield [7]. Following by a step of demethylation reaction [8] by treating **11** with BBr$_3$, curd compound **12** has been obtained and used directly for the next step. Compound **13** has been yielded by treating **12** with trifluoromethanesulfonic anhydride (Tf$_2$O) in 78% yield [9]. Later on, compound **14** has been synthesized after borylation of **13** with pinacolborane in 71% yield [10]. At last, target monomer **1a** has been obtained in 60% yield after treatment of **14** with copper (II) bromide in a seal tube at 120 °C [10,11].

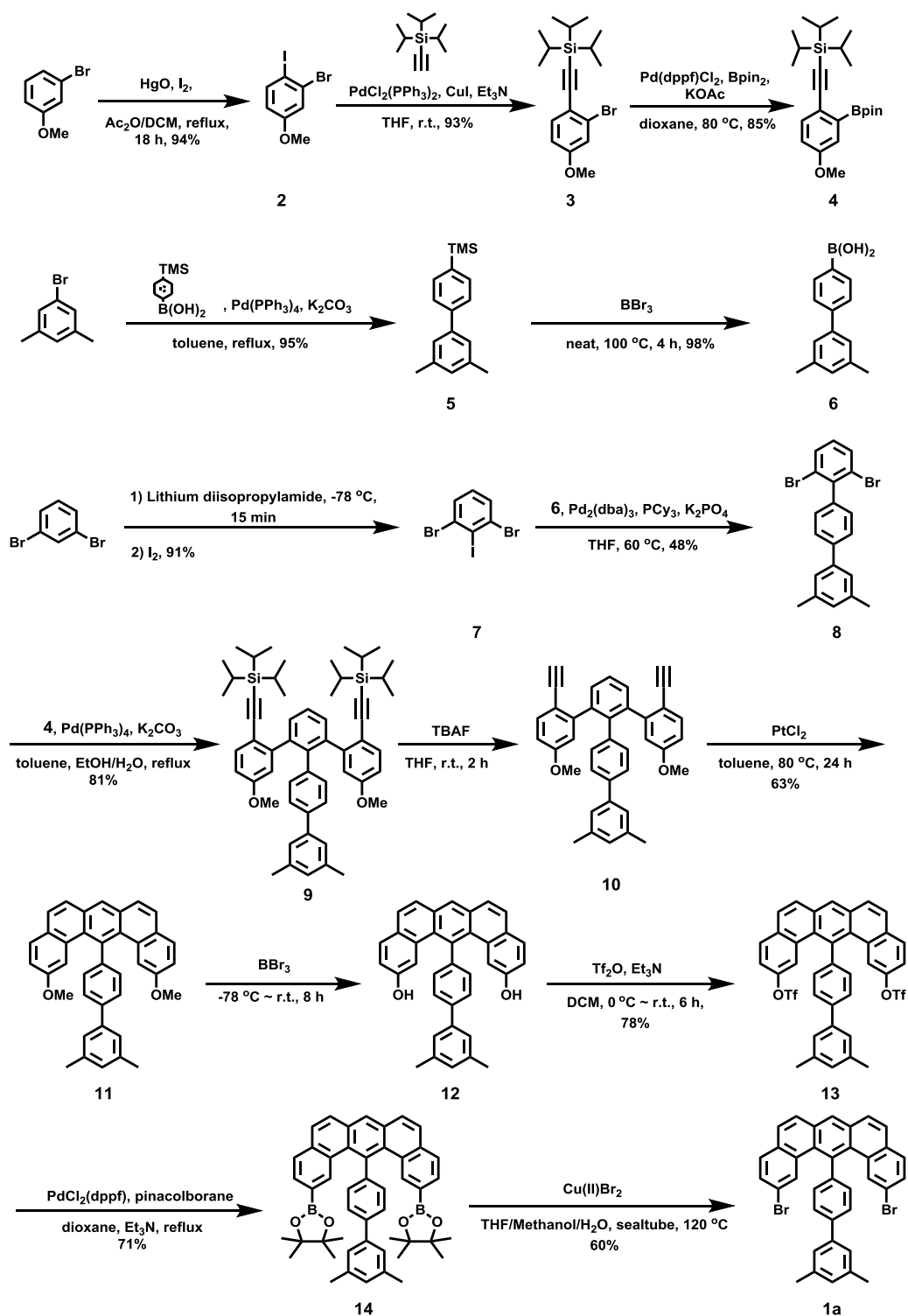

**Supplementary Scheme 1.** Synthetic route towards monomer **1a.**

**Experimental details and description**

<u>**3-Bromo-4-iodoanisole (2)**</u>

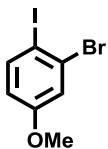

A stirred solution of 3-bromoanisole (10 g, 53.5 mmol), mercury (II) oxide (8.8 g, 40.6 mmol), and acetic anhydride (1 mL) in dichloromethane (100 mL) has been refluxed for 30 min. Then, the iodine (17.6 g, 69.5 mmol) has been added by six portions every 30 min. After refluxing for 12 h and filtration over a pad of celite, the filtrate has been washed with a saturated sodium thiosulfate solution. The aqueous layer has been extracted with dichloromethane (3 x 10 mL), the combined organic layers have been dried over sodium sulfate, and evaporated to remove solvent. Then, purification by flash chromatography (eluent: cyclohexane) has afforded the titled compound **2**.

Colorless oil (Yield = 94%). $^1$H NMR (300 MHz, CD$_2$Cl$_2$) δ 3.77 (s, 3H), 6.62 (dd, *J* = 8.8, 2.9 Hz, 1H), 7.21 (d, *J* = 2.9 Hz, 1H), 7.71 (d, *J* = 8.8 Hz, 1H); $^{13}$C NMR (75 MHz, CD$_2$Cl$_2$) δ 56.19, 89.80, 115.84, 118.96, 130.31, 140.79, 160.85.

**((2-Bromo-4-methoxyphenyl)ethynyl)triisopropylsilane (3)**

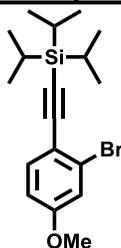

To a mixture of aryl iodide compound **2** (10 g, 32 mmol), bis-(triphenylphosphine)palladium (II) dichloride (448 mg, 0.64 mmol), copper(I) iodide (243.4 mg, 1.28 mmol), triethylamine (14 mL) in tetrahydrofuran (100 mL) has been added drop wise under an argon atmosphere. Next, the liquid compound (triisopropylsilyl) acetylene (8.74 g, 48 mmol) has been added *via* a syringe as well. After the mixture has been stirred at room temperature overnight, diethyl ether (20 mL) has been added to the crude mixture. Then the mixture has been filtered over a short pad of celite to remove catalyst. The organic layer has been washed with brine (5 mL) for three times, and the organic layer has been collected and dried over magnesium sulfate and evaporated. Purification by flash chromatography (eluent: 1% diethyl ether/hexane) has afforded the compound **3**.

Colorless oil (Yield = 93%); $^1$H NMR (300 MHz, CD$_2$Cl$_2$) δ 1.16 (s, 21H), 3.80 (s, 3H), 6.82 (dd, *J* = 8.7, 2.6 Hz, 1H), 7.14 (d, *J* = 2.5 Hz, 1H), 7.44 (d, *J* = 8.6 Hz, 1H); $^{13}$C NMR (75 MHz, CD$_2$Cl$_2$) δ11.98, 19.07, 56.24, 94.56, 105.40, 113.98, 118.33, 118.37, 126.89, 135.07, 160.61; HRMS (MALDI-TOF, positive) m/z calcd for C$_{18}$H$_{27}$BrOSi[M]$^+$ 366.1015, found 366.1078.

**Triisopropyl((4-methoxy-2-(4,4,5,5-tetramethyl-1,3,2-dioxaborolan-2-yl)phenyl)ethynyl)silane (4)**

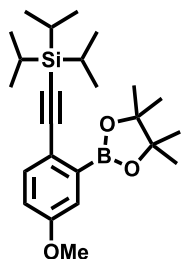

A 250 mL round flask has been charged with compound **3** (19.5 g, 53.1 mmol), bis(pinacolato)diboron (14.8 g, 58.4 mmol), potassium acetate (15.6 g, 159 mmol) and [1,1'-bis(diphenylphosphino)ferrocene]dichloropalladium(II) (1.2 g, 1.6 mmol). Then the stirring mixture has been degassed by argon bubbling for 20 min. Afterwards, the mixture has been stirred overnight at 80 °C under an argon atmosphere. After cooling to room temperature, the mixture has been washed with water and extracted with ethyl acetate (20 mL X 3). The combined organic layer has been washed with brine, dried over magnesium sulfate, and evaporated. At last, the crud residue **4** has been purified by passing through a shot pad of silica gel (eluent: 10% ethyl acetate/hexane) to remove the catalyst and used directly for the next step.

Brown yellow oil (Yield = 85%). $^1$H NMR (300 MHz, CD$_2$Cl$_2$) δ 1.15 (d, $J$ = 1.2 Hz, 23H), 1.33 (s, 12H), 3.81 (s, 3H), 6.89 (dd, $J$ = 8.6, 2.9 Hz, 1H), 7.23 (d, $J$ = 2.8 Hz, 1H), 7.45 (d, $J$ = 8.5 Hz, 1H); $^{13}$C NMR (75 MHz, CD$_2$Cl$_2$) δ 11.86, 18.97, 25.00, 55.63, 84.28, 90.90, 108.06, 116.72, 120.34, 120.80, 135.85, 159.17; HRMS (ESI, positive) m/z calcd for C$_{24}$H$_{40}$BO$_3$Si[M+H]$^+$ 415.2840, found 415.2845.

**(3', 5'-Dimethyl-[1,1'-biphenyl]-4-yl)trimethylsilane (5)**

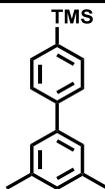

A 250 mL round flask has been charged with 1-bromo-3,5-dimethylbenzene (6 g, 32.4 mmol), (4-(trimethylsilyl)phenyl)boronic acid (9.44 g, 48.6 mmol), potassium carbonate solution (18 g in 10 mL water), ethanol 10 mL, and toluene 50 mL. The mixture has been degassed by argon bubbling for 10 min. Then tetrakis(triphenylphosphino)palladium(0) (1.87 g, 1.62 mmol) has been added. The resulting mixture has been further degassed by argon bubbling for 10 min, and treated with liquid nitrogen bath. After three times freeze-pump-thaw procedure, the mixture has been refluxed overnight. The reaction has been monitored by thin-layer chromatography plate. Once the reaction is completed, the mixture has been washed with deionized water and the aqueous layer has been extracted with ethyl acetate for three times (10 mL x 3). The combined organic layer has been washed with brine, dried over sodium sulfate, and evaporated. The crud product has been purified by silica gel column chromatography (eluent: 5% dichloromethane/hexane) to afford compound **5**.

Colorless oil (Yield = 95%). $^1$H NMR (300 MHz, CD$_2$Cl$_2$) δ 0.32 (s, 9H), 2.39 (s, 6H), 7.02 (m, 1H), 7.24 (ddd, $J$ = 1.9, 1.3, 0.7 Hz, 2H), 7.60 (t, $J$ = 1.3 Hz, 4H); $^{13}$C NMR (75 MHz, CD$_2$Cl$_2$) δ -0.83, 21.72, 125.47, 126.87, 129.55, 134.34, 138.90, 139.65, 141.48, 142.27.

**(3',5'-dimethyl-[1,1'-biphenyl]-4-yl)boronic acid (6)**

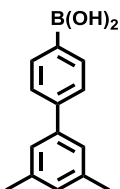

Compound **5** (8 g, 31.4 mmol) has been directly treated with neat boron tribromide (12.6 g, 50.3mmol) under argon atmosphere. A condenser charged with argon has been attached, and the mixture has been heated to 100 °C for 4 h. Once cooled, excess boron tribromide has been distilled off under vacuum at room temperature. The resulting gray-purple solid has been dissolved in dry hexane (50 mL) and cooled to 0 °C with an ice bath. Water has been slowly added drop wise while stirring vigorously until the reaction had been fully quenched. The resulting mixture has been filtered and the white solid has been washed with deionized water and hexane. The white powder has been dried at 80 °C under vacuum overnight, yielding boronic acid **6**, which has been used directly for the next step.

White powder.

**1,3-Dibromo-2-iodobenzene (7)**

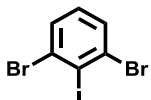

At -75 °C, butyllithium (42.4 mmol) in hexane (50 mL) and diisopropylamine (42.4 mmol) have been added successively to tetrahydrofuran (20 mL). After 15 min, 1, 3-dibromobenzene (5.12 mL, 10 g, 42.4 mmol) has been added. The mixture has been kept at -75 °C for 2 h before a solution of iodine (10.76 g, 42.4 mmol) in tetrahydrofuran (50 mL) is added. After addition of a 10% aqueous solution (0.10 L) of sodium thiosulfate, the mixture has been extracted with diethyl ether for three times (10 mL x 3). The combined organic layer has been washed with water and brine once, and then dried over sodium sulfate before being evaporated to dryness. Upon crystallization from ethanol (100 mL), the colorless platelets have been obtained.

Colorless platelets (Yield = 91%). $^1$H NMR (300 MHz, CD$_2$Cl$_2$) δ 7.10 (t, $J$ = 8.0 Hz, 1H), 7.58 (d, $J$ = 8.1 Hz, 2H); $^{13}$C NMR (75 MHz, CD$_2$Cl$_2$) δ109.67, 131.09, 131.73.

**2,6-Dibromo-3'',5''-dimethyl-1,1':4',1''-terphenyl (8)**

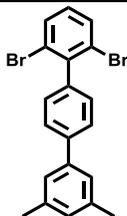

In a glove box, tris(dibenzylideneacetone)dipalladium(0) (1.02 g, 1.11 mmol), tricyclohexylphosphine (1.25 g, 4.45 mmol), compound **7** (8.04 g, 22.23 mmol), and boronic acid **6** (5.03 g, 22.23 mmol) have been added to a reaction vessel that is equipped with a stir bar. The degassed tripotassium phosphate (14.16 g, 66.7 mmol) water solution has been then added, followed by 100 mL anhydrous tetrahydrofuran. The reaction mixture has been then stirred at 60 °C for 3 days. Once the reaction is finished, the reaction mixture has been diluted with ethyl acetate, and then extracted by ethyl acetate for three times. The combined organic layer have been washed three times with water and once with brine, then dried over magnesium sulfate and evaporated. The final product **8** has been obtained after purification by silica gel column chromatography (eluent: 10% dichloromethane/hexane).

Colorless oil (Yield = 48%). Mp: 104.2-104.9 °C; $^1$H NMR (300 MHz, CD$_2$Cl$_2$) δ 2.40 (s, 6H), 7.04 (tt, $J$ = 1.6, 0.8 Hz, 1H), 7.12 (t, $J$ = 8.0 Hz, 1H), 7.23 – 7.35 (m, 4H), 7.64 – 7.73 (m, 4H); $^{13}$C NMR (75 MHz, CD$_2$Cl$_2$) δ 21.72, 125.08, 125.52, 127.31, 129.70, 130.19, 130.56, 132.50, 138.96, 140.58, 140.89, 141.62, 143.25; HRMS (APPI-TOF, positive) m/z calcd for C$_{20}$H$_{16}$Br$_2$ [M]$^+$ 413.9619, found 413.9619.

**2,6-di[5-methoxy-2-((triisopropylsilyl)ethynyl)phenyl]-3'',5''-dimethyl-1,1':4',1''-terphenyl (9)**

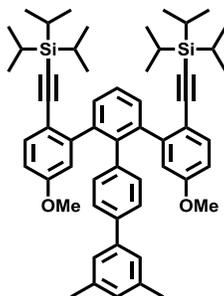

A 100 mL round flask has been filled with compound **8** (1.16 g, 2.79 mmol), boronic ester **4** (3.47 g, 8.36 mmol) in 50 ml toluene and 5 mL potassium carbonate (2.31 g, 16.72 mmol) water solution. After degassed by argon bubbling for 10 min, the tetrakis(triphenylphosphino)palladium(0) (322 mg, 0.28 mmol) has been added. The reaction mixture has been then refluxed overnight, and stopped after thin-layer chromatography indicated that the starting material is totally converted. After cooling down to room temperature, the mixture has been extracted with ethyl acetate for three times (10 mL x 3), and then the combined organic layer has been washed three times with water and dried over magnesium sulfate, then evaporated. The residue has been purified by silica gel column chromatography (eluent: 10% ethyl acetate/hexane) yielding compound **9**.

Yellow solid (Yield = 81%). Mp: 180.8-181.2 °C; $^1$H NMR (300 MHz, CD$_2$Cl$_2$) δ 1.02 (s, 42H), 2.30 (s, 6H), 3.50 (s, 6H), 6.30 (d, $J$ = 2.6 Hz, 2H), 6.63 (dd, $J$ = 8.6, 2.7 Hz, 2H), 7.05 – 7.17 (m, 4H), 7.17 – 7.25 (m, 2H), 7.38 (q, $J$ = 1.9 Hz, 4H), 7.42 (s, 1H); $^{13}$C NMR (75 MHz, CD$_2$Cl$_2$) δ 11.92, 19.04, 21.64, 55.64, 92.66, 107.35, 113.62, 115.91, 116.48, 125.05, 125.93, 126.87, 129.25, 130.72, 131.26, 133.81, 138.72, 139.11, 139.42, 139.46, 140.87, 141.09, 147.54, 159.20; HRMS (ESI, positive) m/z calcd for C$_{56}$H$_{71}$O$_2$Si$_2$[M+H]$^+$ 831.4993, found 831.4979.

**2,6-di[5-methoxy-2-ethynyl-phenyl]-3'',5''-dimethyl-1,1':4',1''-terphenyl (10)**

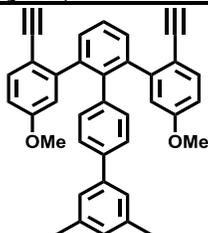

To a solution of compound **9** (1.5 g, 1.8 mmol) in 50 mL of dry tetrahydrofuran, a solution of tetra-*n*-butylammonium fluoride (5.69 g, 18 mmol) in tetrahydrofuran (10 mL) has been added. After stirring at room temperature for 2 h, water has been added to the reaction mixture and tetrahydrofuran has been removed in *vacuo* at 40 °C. The resulting suspension has been extracted three times with ethyl acetate, and the combined organic layers have been washed five times with water, dried over magnesium sulfate, and evaporated. The yielding white solid has been used directly for the next step without further purification.

White solid.

**14-(3',5'-dimethyl-[1,1'-biphenyl]-4-yl)-2,12-dimethoxy-dibenzo[a,j]anthracene (11)**

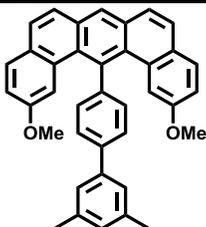

A 100 mL round flask has been charged with compound **10** (1.1 g, 2.12 mmol) and platinum (II) chloride (56.4 mg, 0.21 mmol). Then the mixture has been kept under vacuum conditions for 20 min and refilled with argon. 60 ml of anhydrous toluene has been added by syringe afterwards. The mixture has been heated at 80 °C for 24 h after reaction is completed, as judged by thin-layer chromatography (TLC) plate. The toluene has been then removed under vacuum conditions and the residue has been purified by silica gel column chromatography (eluent: 5% dichloromethane/hexane) yielding the compound **11**.

White solid (Yield = 63%). Mp: 234.4-235.2 °C; $^1$H NMR (300 MHz, CD$_2$Cl$_2$) δ 2.44 (s, 6H), 3.27 (s, 6H), 7.00 (s, 3H), 7.04 (d, $J$ = 2.5 Hz, 1H), 7.10 (s, 1H), 7.31 (dt, $J$ = 1.5, 0.8 Hz, 2H), 7.61 – 7.77 (m, 8H), 7.87 – 7.95 (m, 2H), 8.36 (s, 1H); $^{13}$C NMR (75 MHz, CD$_2$Cl$_2$) δ 21.77, 55.13, 111.74, 117.12, 125.18, 125.37, 128.33, 129.18, 129.92, 130.09, 130.23, 132.45, 132.86, 133.15, 139.34, 144.96, 156.99 ; HRMS (ESI, positive) m/z calcd for C$_{38}$H$_{31}$O$_2$[M+H]$^+$ 519.2324, found 519.2302.

**14-(3',5'-dimethyl-[1,1'-biphenyl]-4-yl)-dibenzo[a,j]anthracene-2,12-diol (12)**

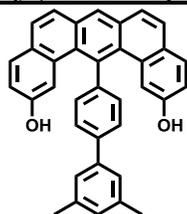

Compound **11** (250 mg, 0.48 mmol) has been dissolved in 40 mL dry dichloromethane under argon atmosphere. Then, 5.8 mL 1M BBr$_3$ (1.45 g, 5.78 mmol in dichloromethane) have been added drop wise to the solution at 0 °C. The solution has been then allowed to warm to room temperature and stirred for 6 h. The reaction has been monitored by thin-layer chromatography plate. Once the reaction is completed, 10 mL water have been slowly added at 0 °C. The mixture has been washed with water and extracted with dichloromethane for three times. Then, the organic layer has been dried over magnesium sulfate, and evaporated. The residue has been re-precipitated from dichloromethane/hexane (1:50) and used directly for the next step without further purification.

Pale green powder.

**14-(3',5'-dimethyl-[1,1'-biphenyl]-4-yl)-dibenzo[a,j]anthracene-2,12-diylbis(trifluoromethanesulfonate) (13)**

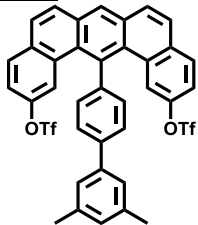

Crud compound **12** (236 mg, 0.48 mmol) has been dissolved in 20 mL dichloromethane and cooled to 0 °C. Then 0.36 mL triethylamine (2.6 mmol) have been added drop wise. Afterwards, 1 M Tf$_2$O (1.44 mL) solution has been added by syringe. The mixture solution has been allowed to warm to room temperature and stirred for 4 h. Once thin-layer chromatography plate shows the complete of the reaction, the dichloromethane has been removed and the residue has been purified by silica gel column chromatography (eluent: 10% ethyl acetate/hexane) yielding compound **13**.

White solid (Yield = 78%). Mp: 305.9-306.7 °C; $^1$H NMR (300 MHz, CD$_2$Cl$_2$) δ 2.45 (s, 6H), 7.11 (s, 1H), 7.30 (d, $J$ = 2.5 Hz, 2H), 7.38 (dd, $J$ = 8.7, 2.5 Hz, 2H), 7.43 (s, 2H), 7.53 (d, $J$ = 8.2 Hz, 2H), 7.79 (s, 1H), 7.82 (s, 1H), 7.92 (d, $J$ = 2.8 Hz, 2H), 7.94 – 7.98 (m, 3H), 8.00 (s, 1H), 8.51 (s, 1H); $^{13}$C NMR (75 MHz, CD$_2$Cl$_2$) δ 21.49, 116.39, 119.74, 122.06, 125.89, 127.94, 128.04, 128.83, 129.15, 129.64, 130.65, 130.87, 131.01, 132.46, 132.78, 134.23, 138.64, 139.91, 140.83, 142.22, 143.72, 146.60; HRMS (ESI, positive) m/z calcd for C$_{38}$H$_{25}$F$_6$O$_6$S$_2$[M+H]$^+$ 755.0997, found 755.0993.

**14-(3',5'-dimethyl-[1,1'-biphenyl]-4-yl)-dibenzo[a,j]anthracene-2,12-diyl-bis(4,4,5,5-tetramethyl-1,3,2-dioxaborolane) (14)**

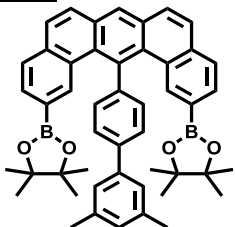

A Schlenk tube has been charged with compound **13** (110 mg, 0.145 mmol), [1,1'-bis(diphenylphosphino)ferrocene]dichloropalladium(II) (6 mg, 0.007 mmol), 5mL dry dioxane and triethylamine (0.12 mL, 0.87 mmol). The solution has been degassed by argon bubbling for 10 min and pinacolborane (0.08 mL, 0.58 mmol) has been added. The mixture has been then heated at refluxing temperature for 12 h. Then the solvent has been removed under vacuum and the residue purified by silica gel chromatography (eluent: 5 % ethyl acetate/hexane) to afford compound **14**.

Light yellow solid (Yield = 71%). Mp: 282.6-283.2 °C; $^1$H NMR (300 MHz, CD$_2$Cl$_2$) δ 1.02 (24H, d, $J$ = 7.3 Hz), 2.40 (6H, d, $J$ = 10.1 Hz), 7.01 (1H, s), 7.48 – 7.54 (4H, m), 7.69 – 7.73 (4H, m), 7.78 – 7.81 (2H, m), 7.85 – 7.88 (2H, d, $J$ = 8.8 Hz), 7.96 – 7.99 (4H, m), 8.35 (1H, s); $^{13}$C NMR (75 MHz, CD$_2$Cl$_2$) δ 144.05, 140.81, 140.42, 139.18, 138.46, 136.76, 136.58, 132.03, 131.83, 131.18, 129.34, 129.17, 128.95, 128.53, 128.48, 128.01, 125.40, 83.95, 54.87, 54.43, 54.00, 53.57, 53.13, 24.77, 21.69; HRMS (APPI-TOF, positive) m/z calcd for C$_{48}$H$_{48}$B$_2$O$_4$[M]$^+$ 710.3739, found 710.3778.

**2,12-dibromo-14-(3',5'-dimethyl-[1,1'-biphenyl]-4-yl)-dibenzo[a,j]anthracene (1a)**

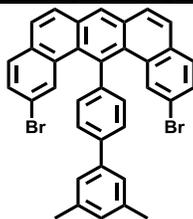

A 12 mL sealtube has been charged with compound **14** (50 mg, 0.07 mmol), copper (II) bromide (95 mg, 0.42 mmol), 2 mL tetrahydrofuran, 6 mL methanol and 4 mL water. The tube has been degassed by argon bubbling for 10 min, then sealed and heated at 120 °C overnight. After cooling to the room temperature, the mixture has been extracted with dichloromethane for three times (5 mL x 3). The combined organic layers have been dried over magnesium sulfate, and then evaporated. The residue has been purified by silica gel chromatography (eluent: 10 % dichloromethane/hexane) and re-precipitated from dichloromethane/methanol (1:10) to yield final product **1a**.

Colorless solid (Yield = 60%). Mp: 267.9-268.5 °C; $^1$H NMR (300 MHz, CD$_2$Cl$_2$) δ 2.37 (s, 6H), 7.02 (s, 1H), 7.37 – 7.49 (m, 8H), 7.59 – 7.67 (m, 4H), 7.78 – 7.90 (m, 4H), 8.35 (s, 1H); $^{13}$C NMR (75 MHz, CD$_2$Cl$_2$) δ 21.79, 119.77, 125.96, 127.76, 128.04, 128.39, 128.73, 129.63, 129.83, 130.33, 130.49, 131.27, 132.54, 132.71, 133.10, 133.38, 139.05, 143.46; HRMS (APPI-TOF, positive) m/z calcd for C$_{36}$H$_{24}$Br$_2$[M]$^+$ 614.0245, found 614.0246.

**C. Synthetic route to monomer 1b**

"U-shaped" monomer **1b** has been synthesized based on the pyrylium chemistry, which is presented in Supplementary Scheme 2. First, compound **15** has been prepared by condensation of 2-bromo-7-hydroxynaphthalene and corresponding aryl aldehyde (R = 3, 5-dimethylphenyl) under neat condition (*1*), followed by oxidation with lead (IV) oxide to afford compound **16**. Then, crud product of compound **16** has been directly treated with tetrafluoroboric acid solution (48 wt. % in water) to afford pyrylium salt **17** (*2, 3*). Finally, after a condensation reaction of **17** with sodium 2-phenylacetate (*4, 5*), target monomer **1b** has been obtained in 38% yield.

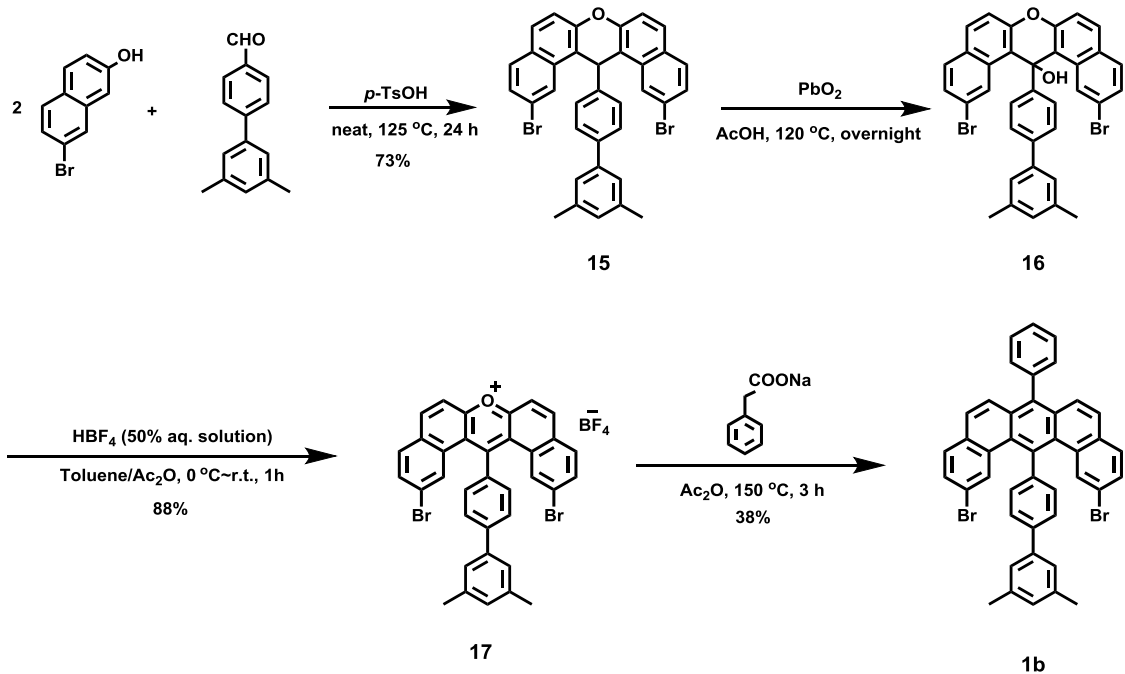

**Supplementary Scheme 2.** General synthetic route towards monomer **1b**.

**Experimental details and description**

**2,12-Dibromo-14-(3',5'-dimethyl-[1,1'-biphenyl]-4-yl)-14H-dibenzo[a,j]xanthene (15)** :

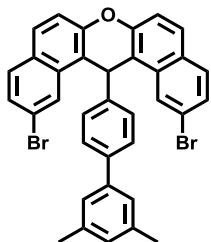

To a mixture of 3',5'-dimethyl-[1,1'-biphenyl]-4-carbaldehyde (2.36 g, 11.2 mmol) and 7-bromo-naphthol (5g, 22.4 mmol), *p*-toluenesulfonic acid monohydrate (*p*-TSA) (43 mg, 0.22 mmol) has been added. The reaction mixture has been stirred magnetically at 125 °C for 24 h. The reaction has been monitored by thin-layer chromatography (TLC) plate. Once the reaction is completed, the mixture is washed with EtOH–H$_2$O (1:3). The crude product has been purified by recrystallization from ethanol to afford target compound **15**.

White needle (Yield = 73%); Mp: < 319 °C; $^1$H NMR (500 MHz, C$_2$D$_2$Cl$_4$) δ 6.28 (s, 1H), 6.93 (s, 1H), 7.05 (s, 2H), 7.43 (d, *J* = 7.4 Hz, 2H), 7.53 (m, 6H), 7.77 (dd, *J* = 35.5, 8.7 Hz, 4H), 8.54 (s, 2H); $^{13}$C NMR (125 MHz, C$_2$D$_2$Cl$_4$) δ 21.27, 37.51, 37.51, 73.56, 73.78, 73.78, 74.00, 115.91, 118.46, 121.47, 124.82, 125.01, 127.42, 127.72, 128.21, 128.88, 128.95, 129.30, 130.36, 132.33, 138.08, 139.53, 139.91, 142.84, 148.98; HRMS (MALDI-TOF, positive) *m/z* calcd for 620.0173, found 620.0178.

### 2,12-Dibromo-14-(3',5'-dimethyl-[1,1'-biphenyl]-4-yl)-14H-dibenzo[a,j]xanthen-14-ol (16)

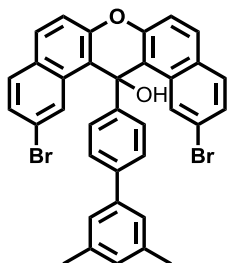

Compound **15** (6.6 g, 10.6 mmol) and lead dioxide (PbO$_2$) (3.8 g, 15.9 mmol) in glacial acetic acid (50 mL) have been stirred while heating on an oil bath at 120 °C for 12 h. The cooled mixture has been poured into crushed ice and the solid residue **16** (crude compound) has been filtered off, dried under vacuum at 80 °C and used directly for the next step.

### 2,12-Dibromo-14-(3',5'-dimethyl-[1,1'-biphenyl]-4-yl)-14-dibenzo[a,j]xanthenylium tetrafluoroborate (17)

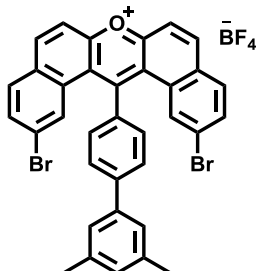

Compound **16** (2 g, 3.1 mmol) in acetic anhydride (15 ml) and toluene (10 mL) has been cooled to 0 °C and treated with tetrafluoroboric acid solution (48 wt. % in water) (3.91 mL, ca. 31 mmol) until no further precipitation occurred. The cooled solution has been filtered and washed with anhydrous ether to yield **17**.

Orange red powder (yield=88%). $^1$H NMR (500 MHz, C$_2$D$_2$Cl$_4$) δ 2.50 (s, 6H), 7.20 (s, 1H), 7.46 (s, 2H), 7.52 (s, 2H), 7.61 (d, $J$ = 7.8 Hz, 2H), 7.95 (d, $J$ = 8.3 Hz, 2H), 8.08 (d, $J$ = 8.4 Hz, 2H), 8.20 (d, $J$ = 7.8 Hz, 2H), 8.32 (d, $J$ = 9.0 Hz, 2H), 8.81 (d, $J$ = 9.1 Hz, 2H); $^{13}$C NMR (125 MHz, C$_2$D$_2$Cl$_4$) δ 21.47, 73.56, 73.78, 74.00, 117.54, 121.13, 125.45, 126.18, 126.58, 129.74, 130.42, 130.93, 131.30, 131.45, 132.59, 133.07, 135.37, 138.92, 139.37, 146.37, 146.90, 159.45, 167.67; HRMS (MALDI-TOF, positive) *m/z* calcd for C$_{35}$H$_{23}$Br$_2$O$^+$[M]$^+$ 617.0116, found 617.0127.

### 2,12-Dibromo-14-(3',5'-dimethyl-[1,1'-biphenyl]-4-yl)-7-phenyl-dibenzo[a,j]anthracene (1b)

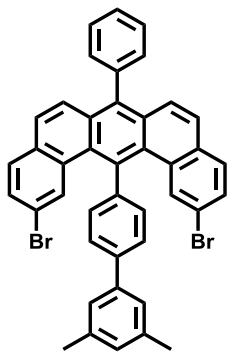

A mixture of the pyrylium salt **17** (4.2g, 5.9 mmol) and sodium 2-phenylacetate (2.8 g, 17.8 mmol) in acetic anhydride (Ac$_2$O, 50 mL) has been stirred at 150 °C for 12 h under argon atmosphere. After cooling to the room temperature, the precipitate has been filtered off and washed with acetic anhydride, then methanol. The crud product has been recrystallized from chloroform and hexane to give target monomer **1b**.

Brown powder (yield= 38%). Mp: > 400 °C; $^1$H NMR (500 MHz, C$_2$D$_2$Cl$_4$) δ 2.48 (s, 6H), 7.12 (s, 1H), 7.48 (d, *J* = 7.3 Hz, 6H), 7.56 – 7.50 (m, 6H), 7.57 (s, 2H), 7.70 – 7.60 (m, 5H), 7.91 (d, *J* = 8.0 Hz, 2H); $^{13}$C NMR (125 MHz, C$_2$D$_2$Cl$_4$) δ 21.48, 73.56, 73.78, 74.00, 118.57, 120.18, 125.62, 125.80, 126.85, 126.97, 127.79, 128.53, 129.04, 129.11, 129.73, 130.30, 131.19, 131.71, 132.00, 132.10, 132.60, 137.39, 137.77, 138.33, 138.83, 141.17, 142.30, 142.67; HRMS (MALDI-TOF, positive) m/z calcd for C$_{42}$H$_{28}$Br$_2$[M]$^+$ 690.0558, found 690.0561.

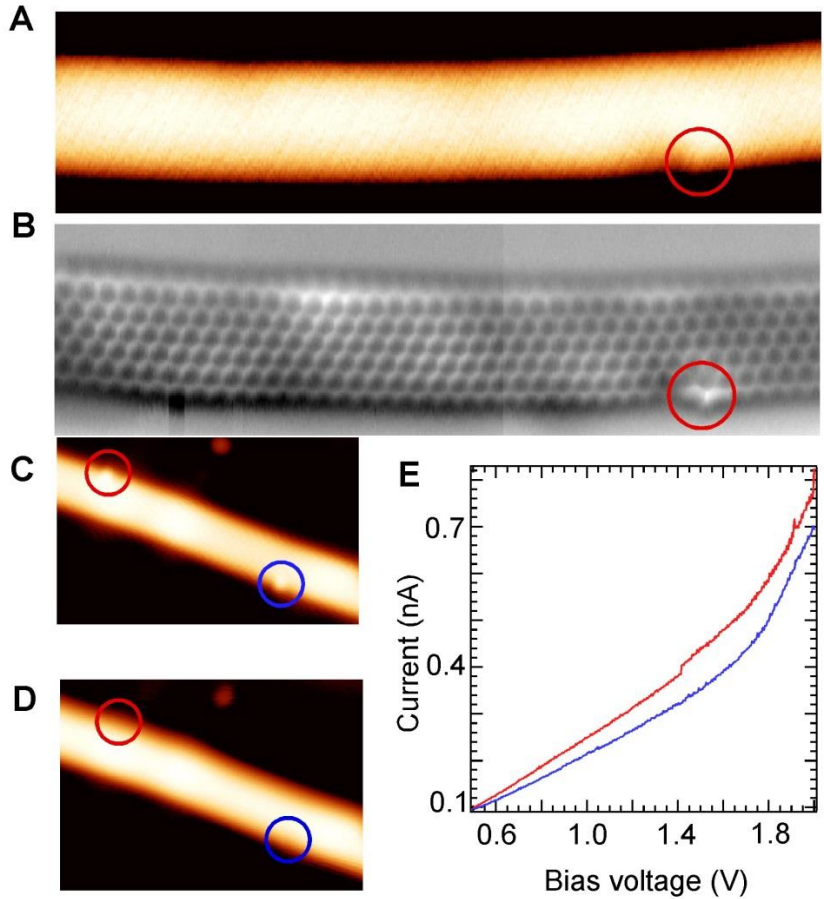

**Fig. S1.**

$H_2$ defects of 6-ZGNRs. (**A**) Constant-current STM image (9.5 nm x 2.8 nm, T = 5 K, U = -1.0 V, I = 20 pA) of 6-ZGNR showing a $H_2$ defect marked by red circles. (**B**) Constant-height AFM frequency-shift image (9.5 nm x 3 nm, oscillation amplitude A = 0.7 Å, sample voltage V = 10 mV) taken at the same place as in (A). (**C**) STM image (10 nm x 5 nm, T = 5 K, U = -1.0 V, I = 20 pA) of 6-ZGNR showing two $H_2$ defects. (**D**) STM image (10 nm x 5 nm, T = 5 K, U = -1.0 V, I = 20 pA) showing the same ribbon in (C) after tip-induced dehydrogenation. (**E**) Current-voltage spectra taken above $H_2$ defects to remove one hydrogen.

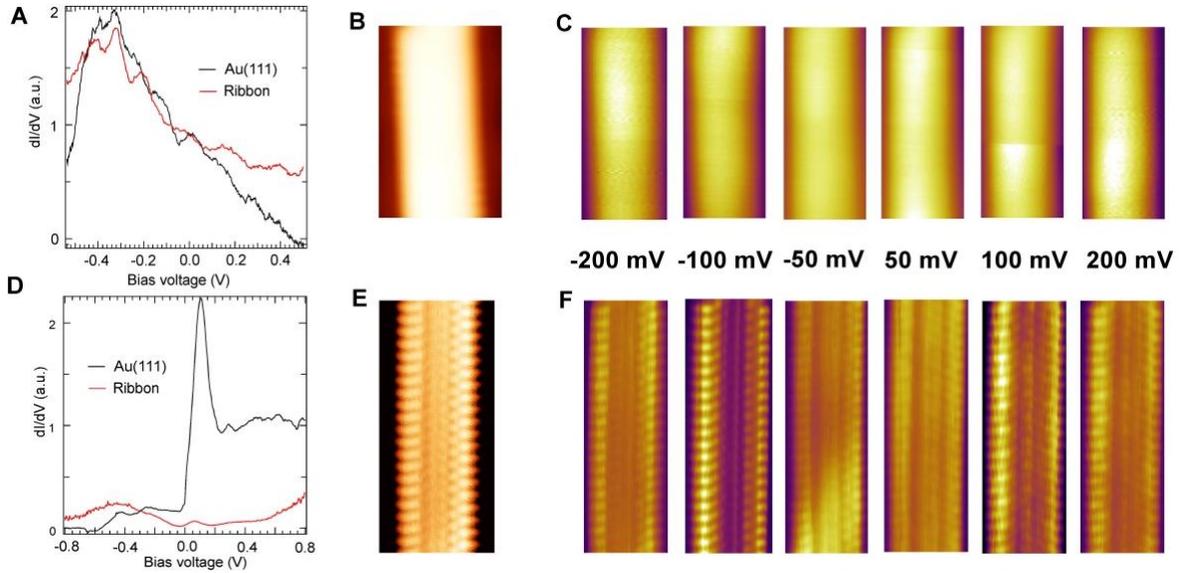

**Fig. S2.**
dI/dV spectra and maps of 6-ZGNR taken with a metallic and a 'special' tip, respectively. (**A**) Differential conductance (dI/dV) spectra taken at the zigzag edge and on the nearby Au(111) substrate using a clean metallic tip. (**B**) Constant-current STM image (2.4 nm x 3.8 nm, T = 5 K, U = -50 mV, I = 1 nA) of 6-ZGNR. (**C**) Constant-height dI/dV maps (1.7 nm x 3.8 nm, T = 5 K) of 6-ZGNR near Fermi level using normal metal tip. No evidence for increased intensity at the zigzag edges is obtained. (**D**) Differential conductance (dI/dV) spectra taken at the zigzag edge and nearby Au(111) using a 'special' tip of unknown termination. (**E**) Constant-current STM image (2.2 nm x 4.7 nm, T = 5 K, U = -0.2 V, I = 1 nA) of 6-ZGNR. (**F**) Constant-height dI/dV maps (1.6 nm x 4.7 nm, T = 5 K) of 6-ZGNR near Fermi level using the 'special' tip showing an increased intensity at zigzag edges.

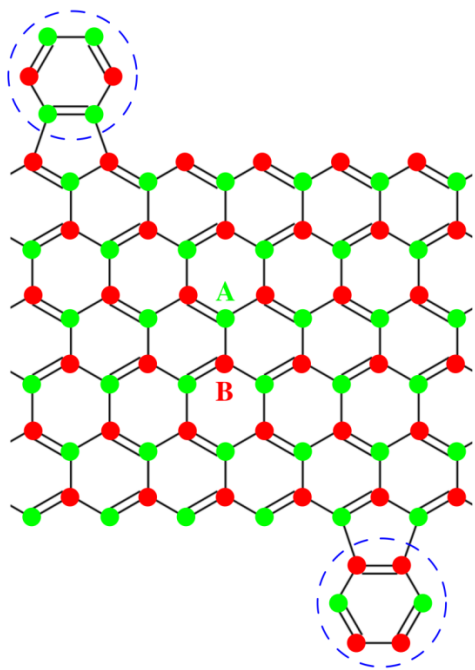

**Fig. S3.**
Geometry frustration at the zigzag backbone. Sub-lattice description of graphene, illustrated for the case of the edge-modified 6-ZGNR. Each atom on the A sub-lattice is surrounded by three B sub-lattice atoms and vice-versa. The blue dotted-circles indicate frustrated arrangement of sublattices at the zigzag backbone forming the pentagonal (defect) sites.

# $^1$H and $^{13}$C NMR spectra

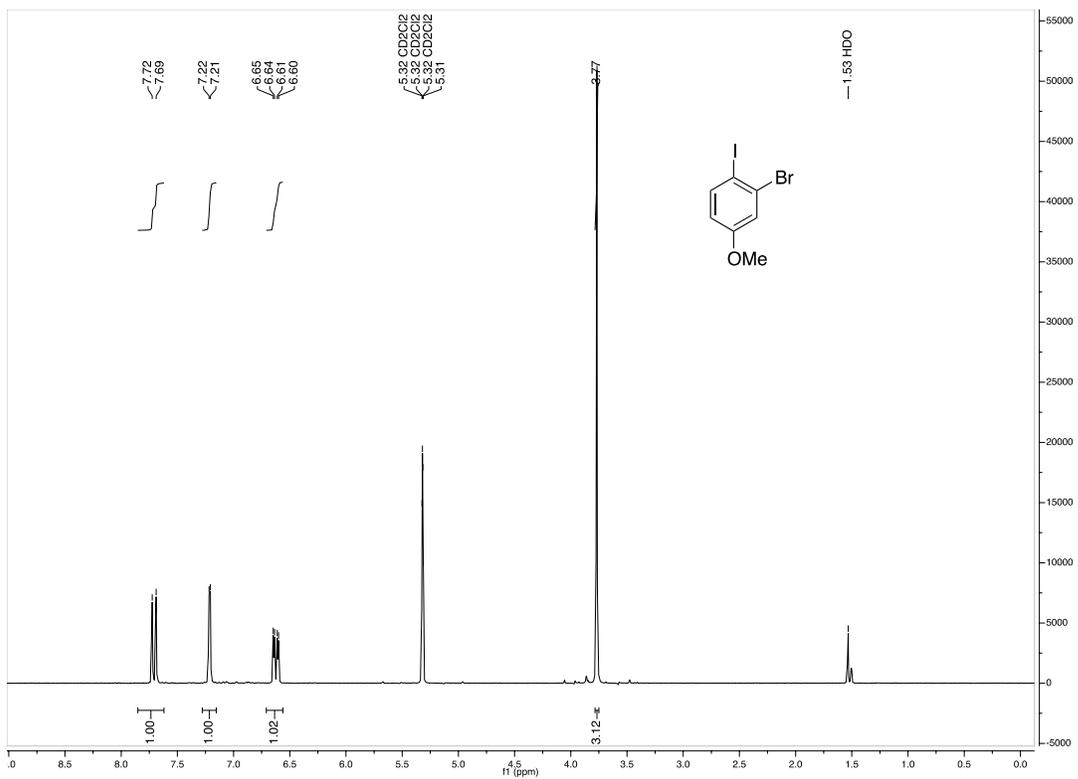

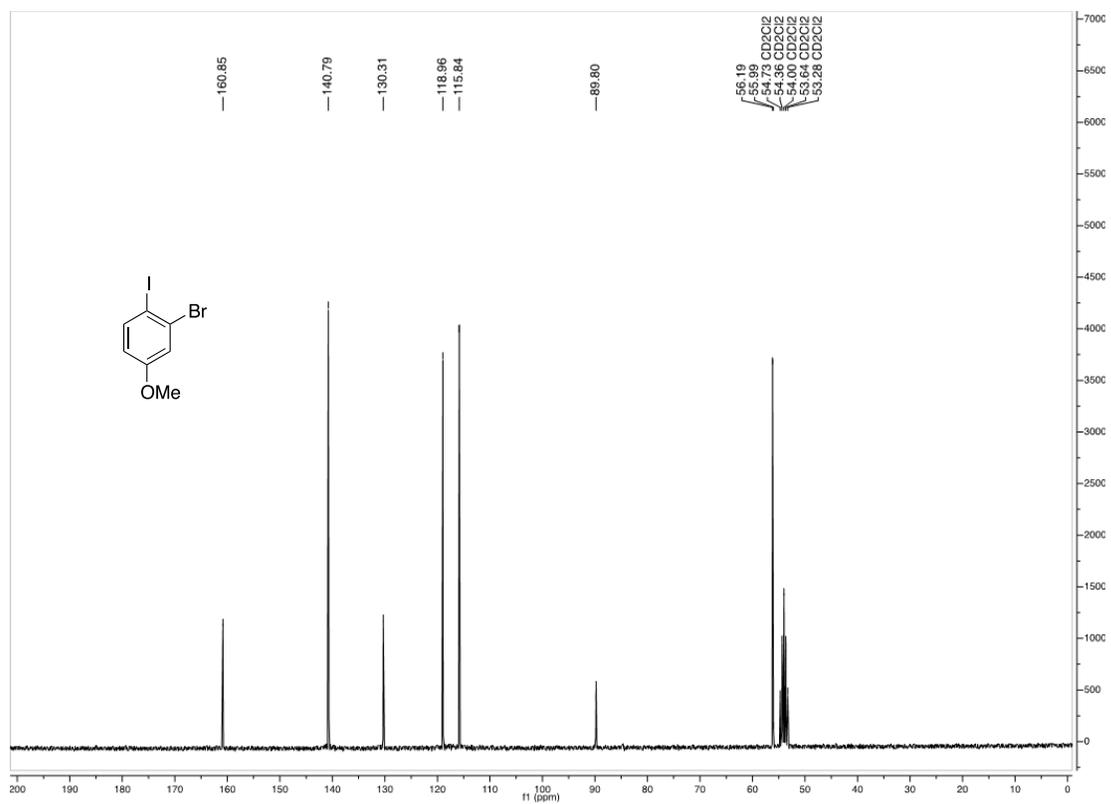
18

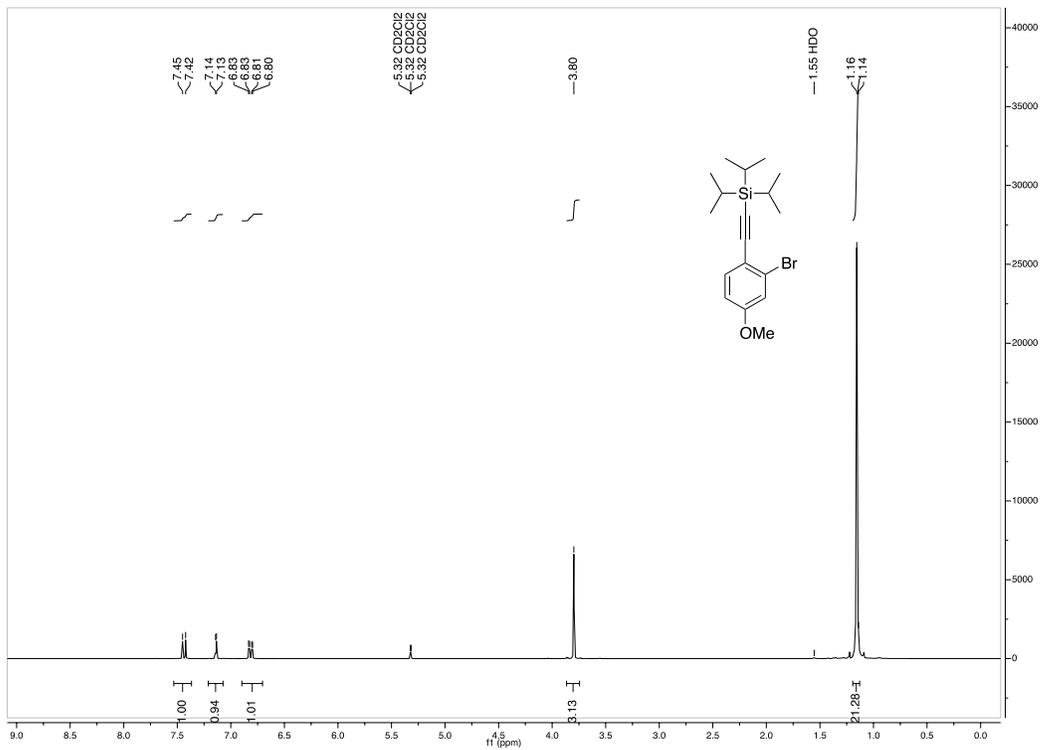
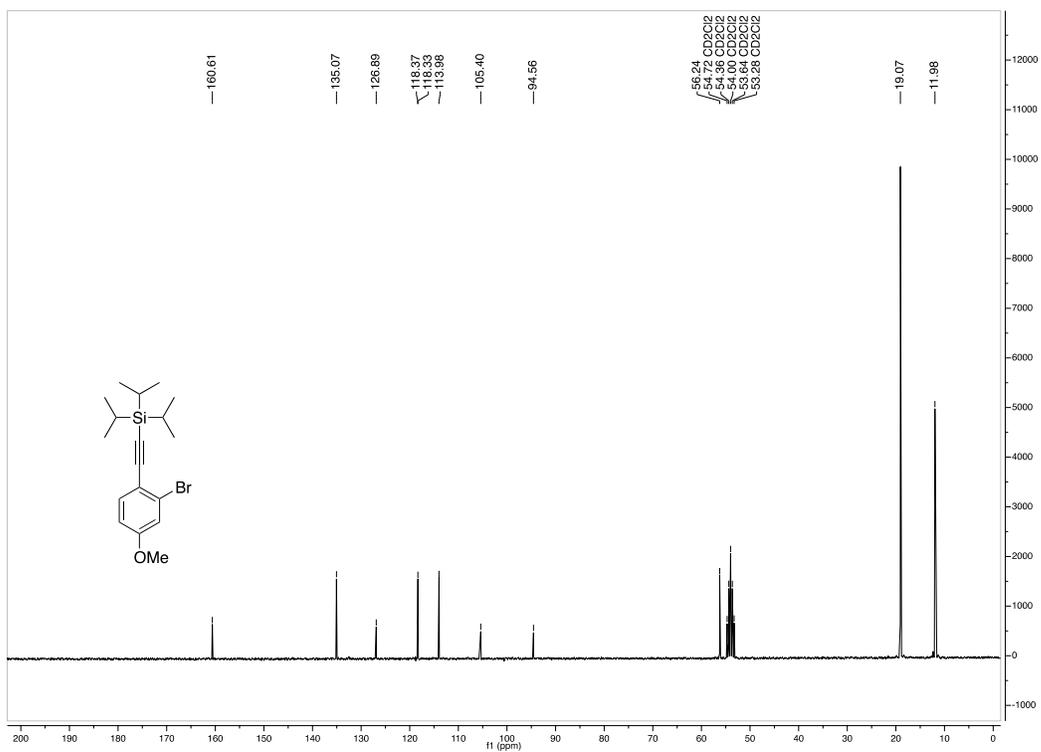

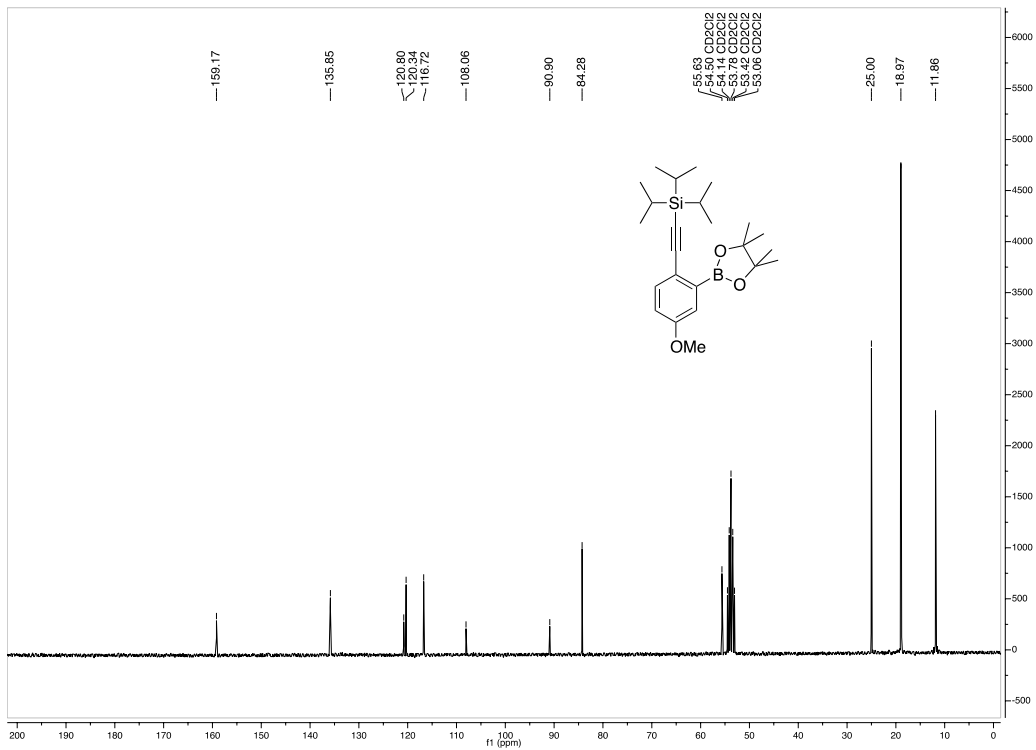
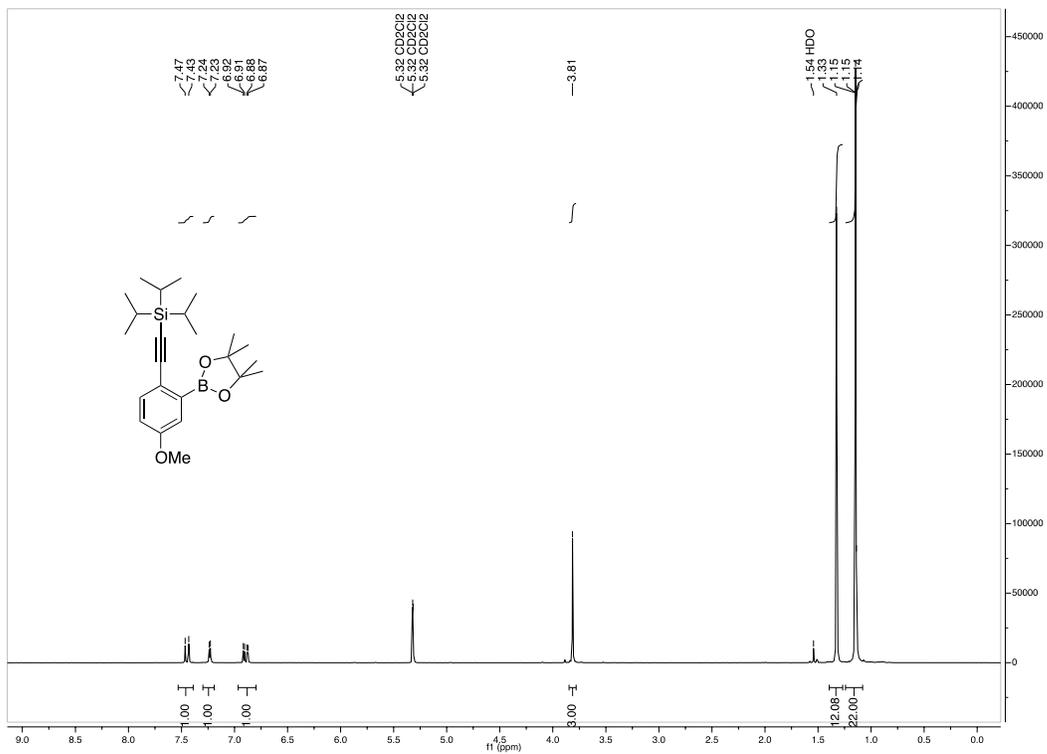

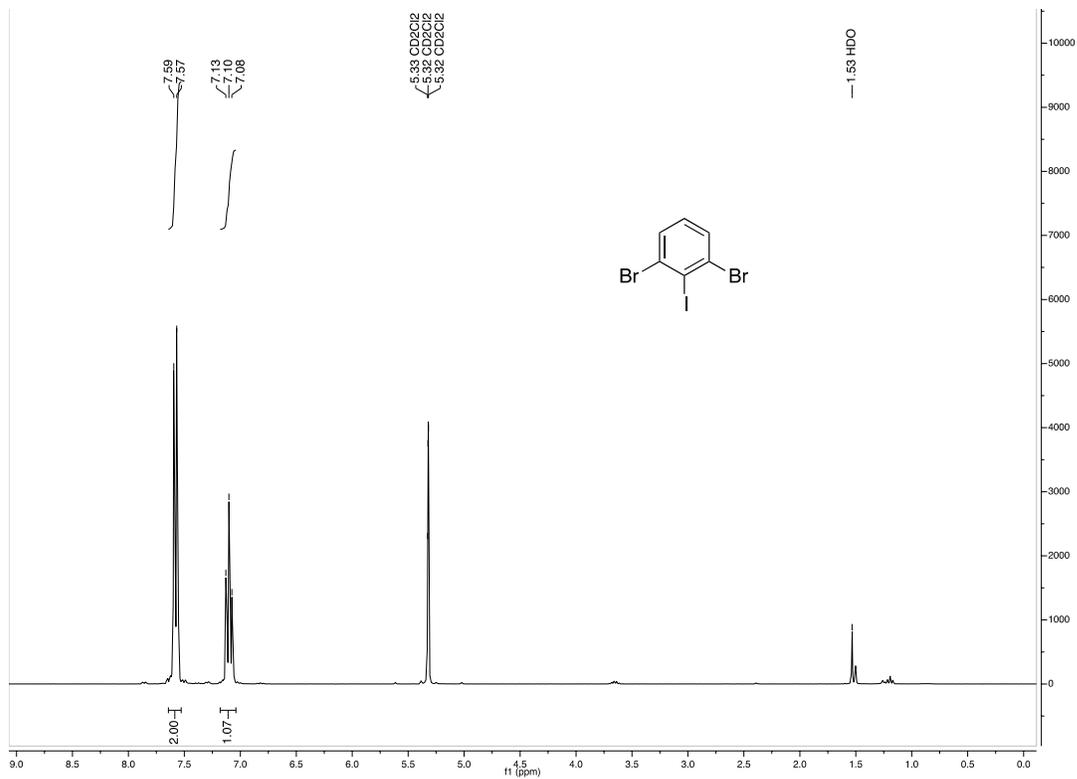

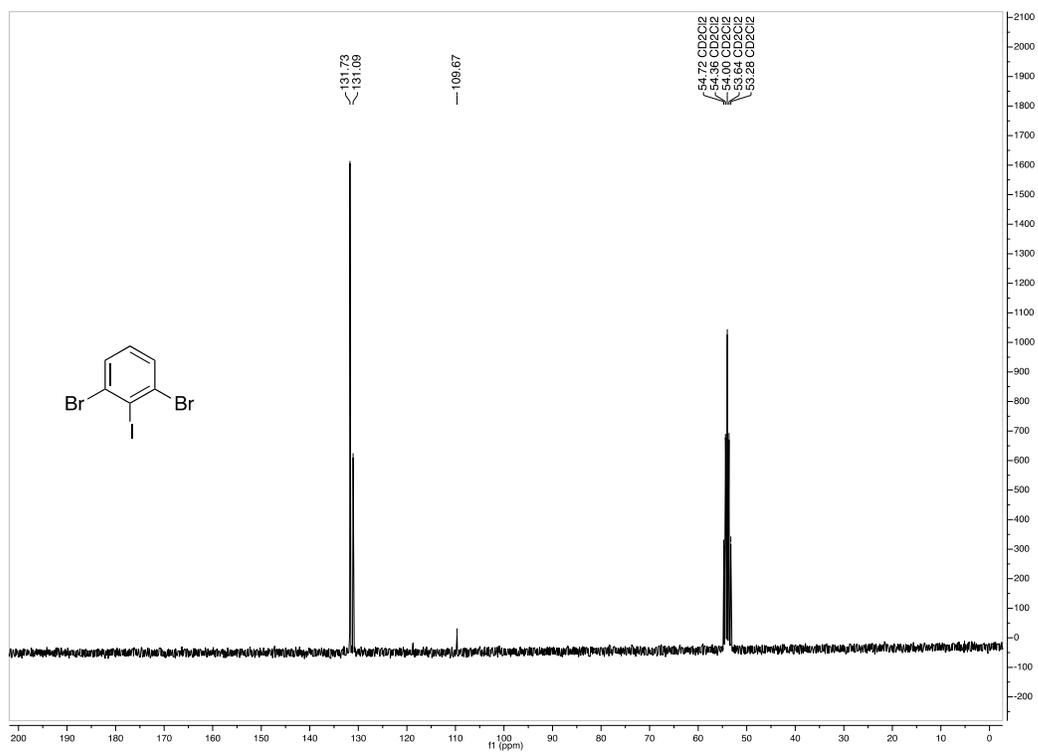

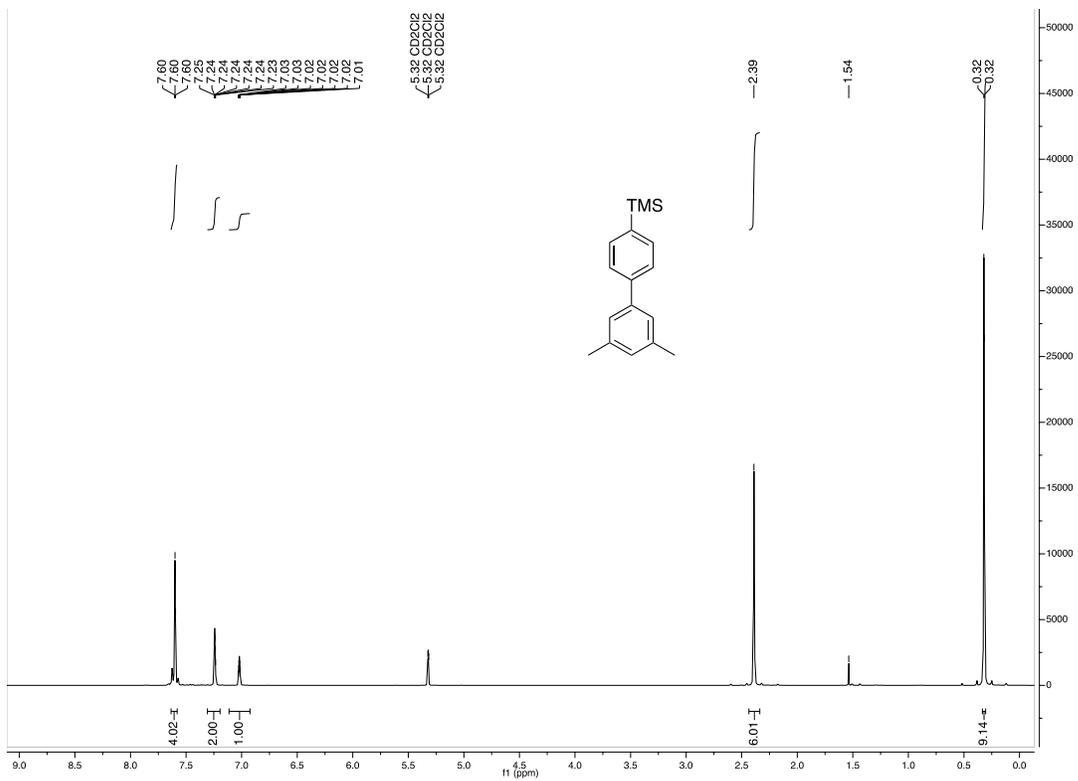
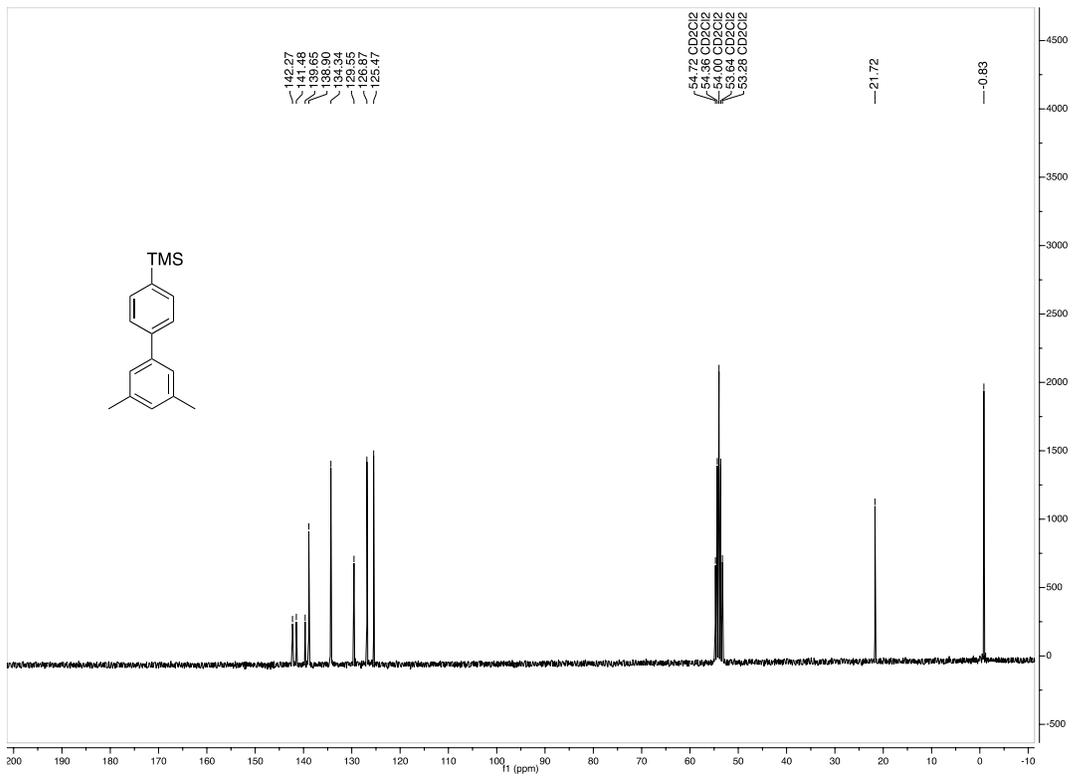

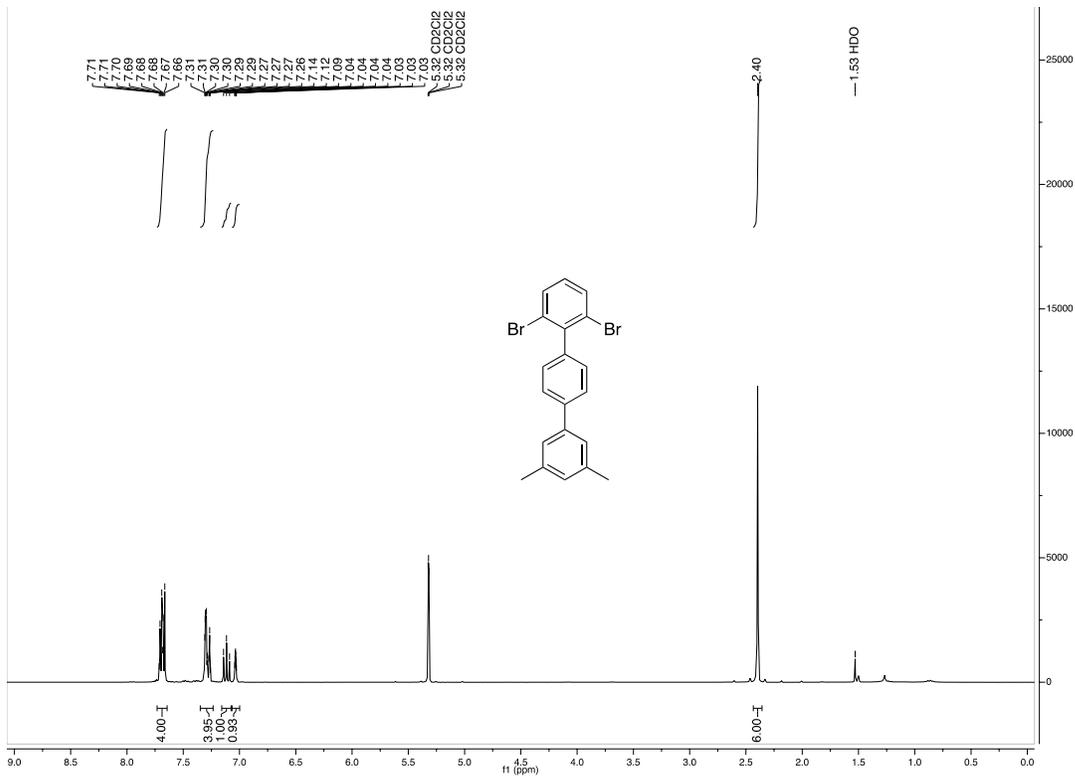
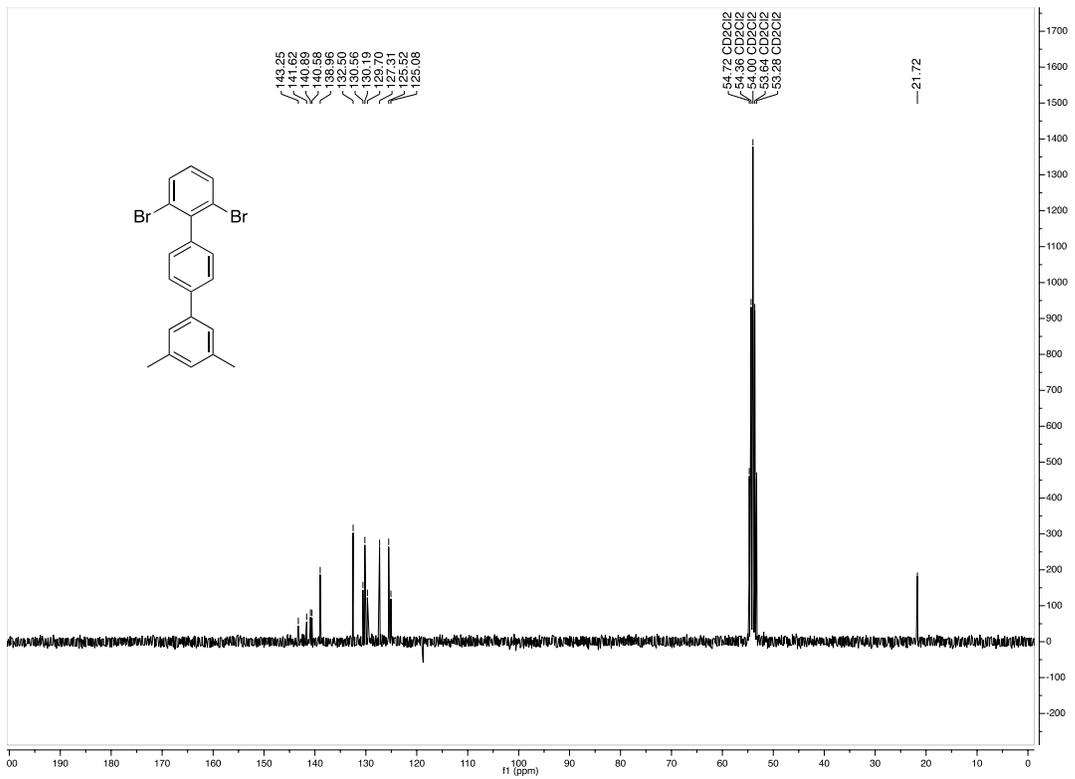

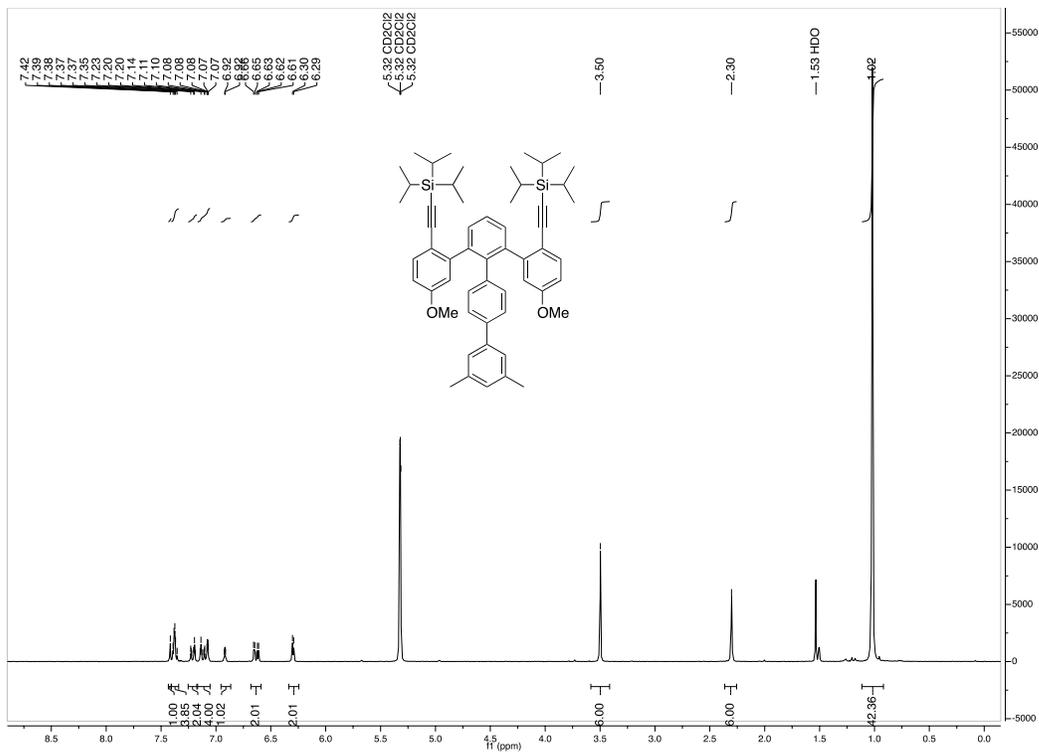

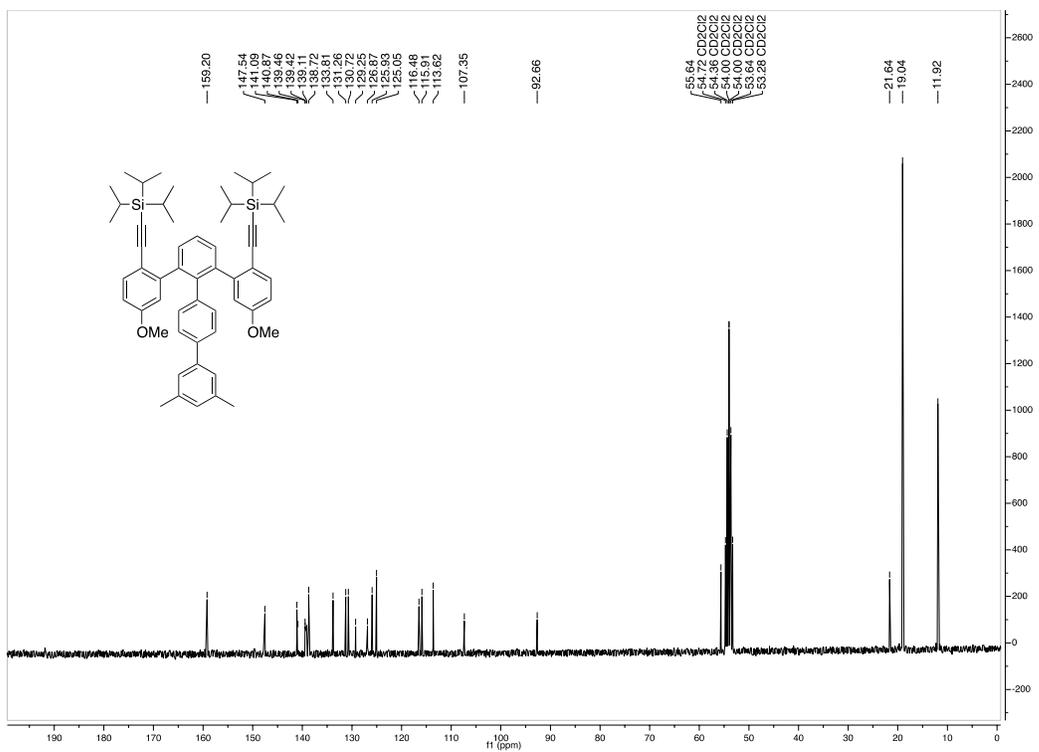

NMR spectra (¹H top, ¹³C bottom) in CD₂Cl₂.

¹H NMR peaks: 8.36, 7.93, 7.92, 7.90, 7.89, 7.75, 7.73, 7.71, 7.70, 7.69, 7.68, 7.66, 7.65, 7.31, 7.30, 7.10, 7.04, 7.03, 5.32 (CD₂Cl₂), 3.27, 2.44, 1.53 (HDO).

Integrations: 1.00, 2.04, 8.19, 2.11, 1.01, 1.06, 3.03, 6.12, 6.22.

¹³C NMR peaks: 156.80, 144.78, 139.16, 132.68, 130.04, 129.91, 129.00, 128.15, 125.18, 126.04 (126.94), 111.56, 54.95, 21.58.

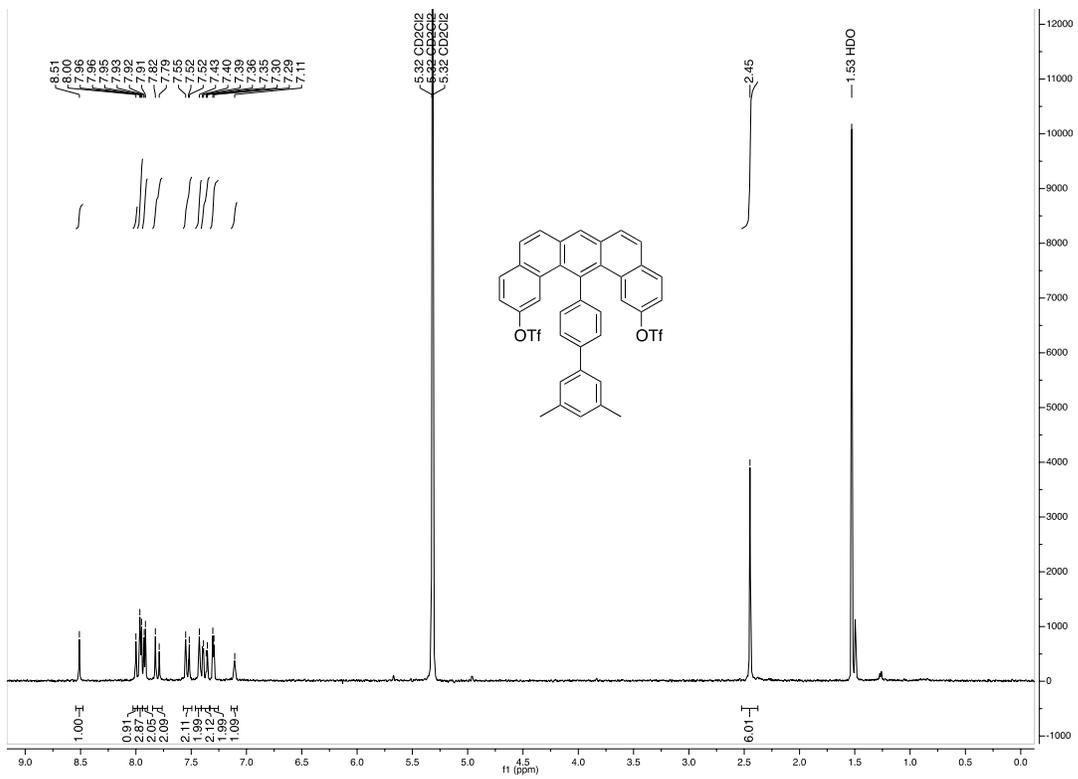

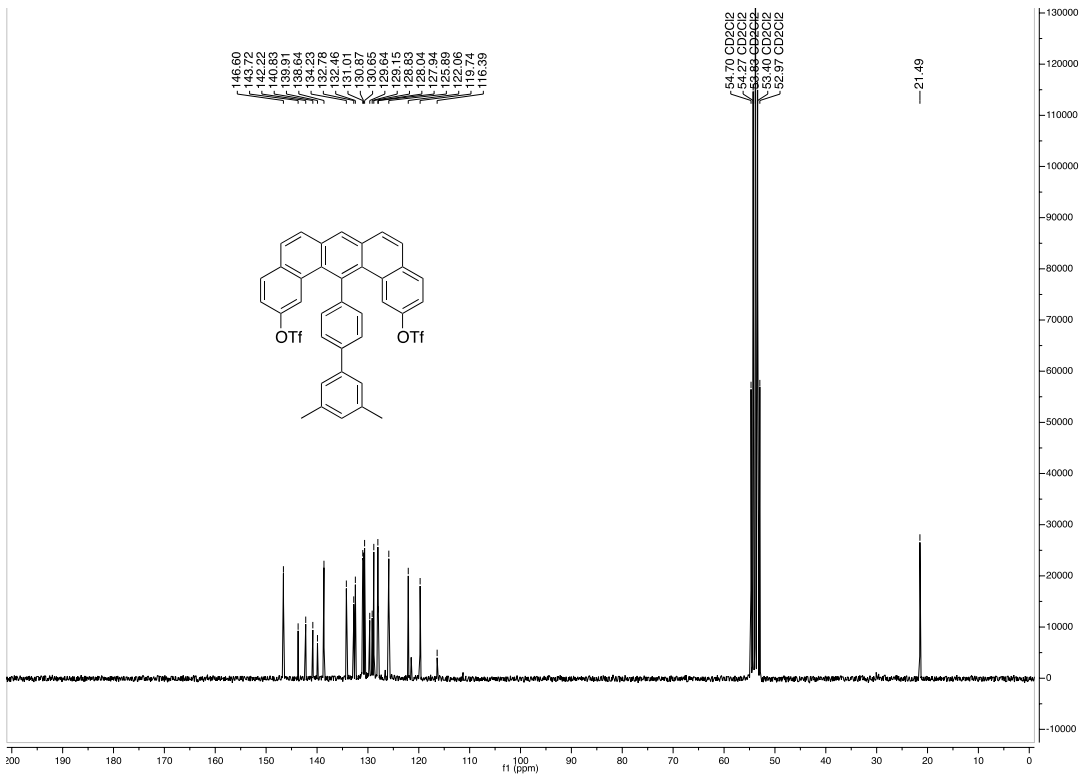

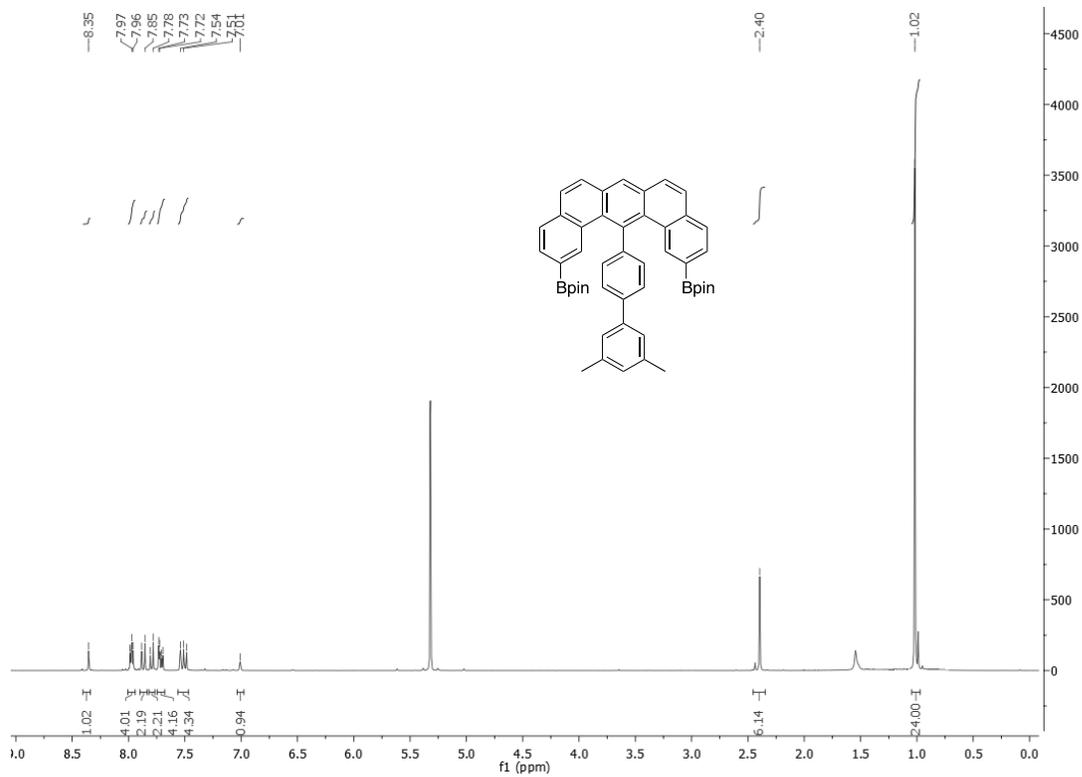
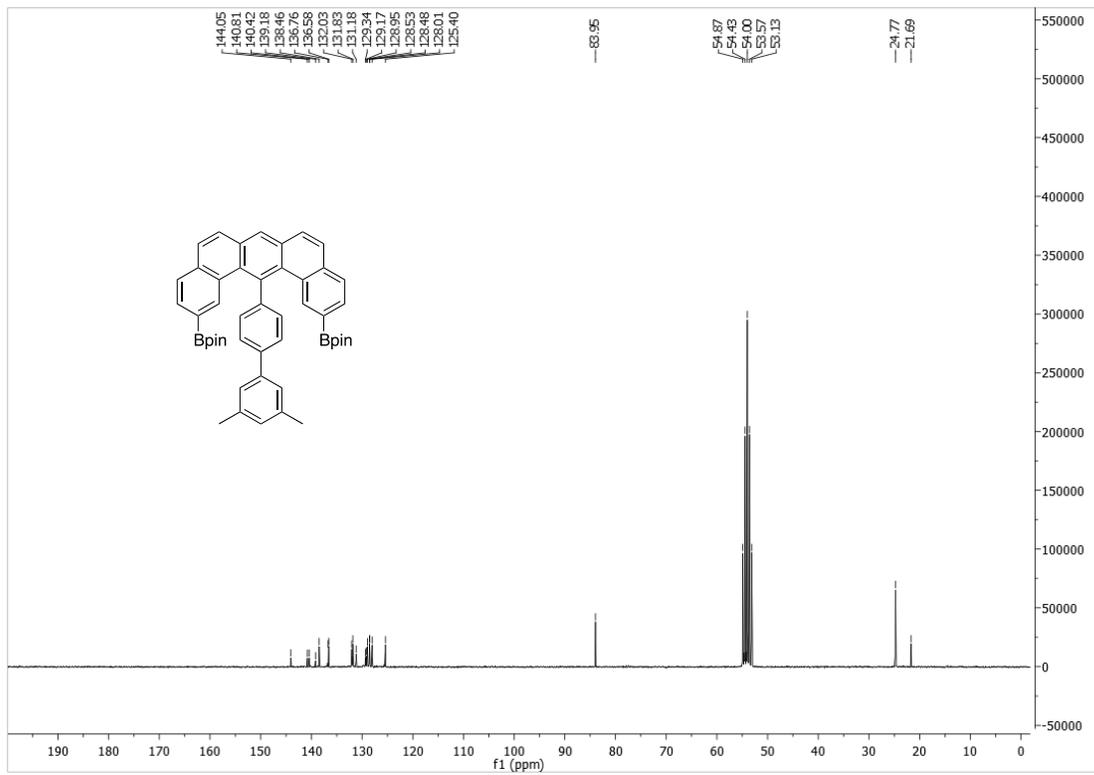

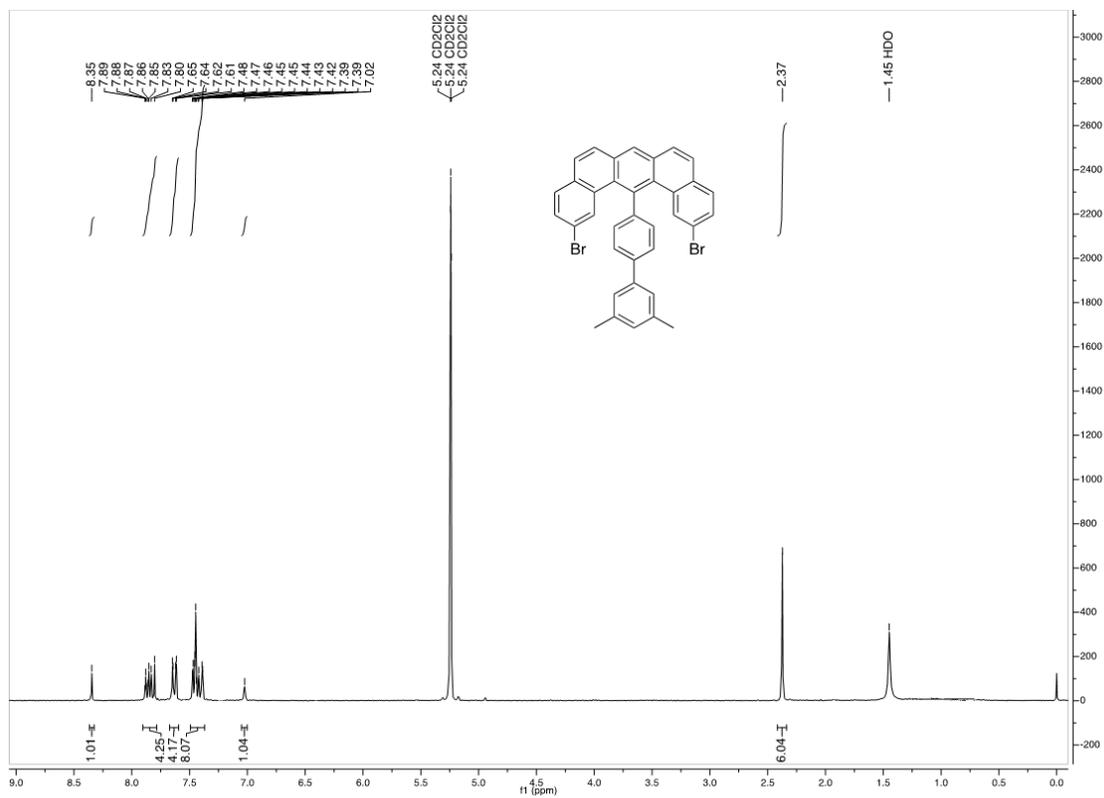

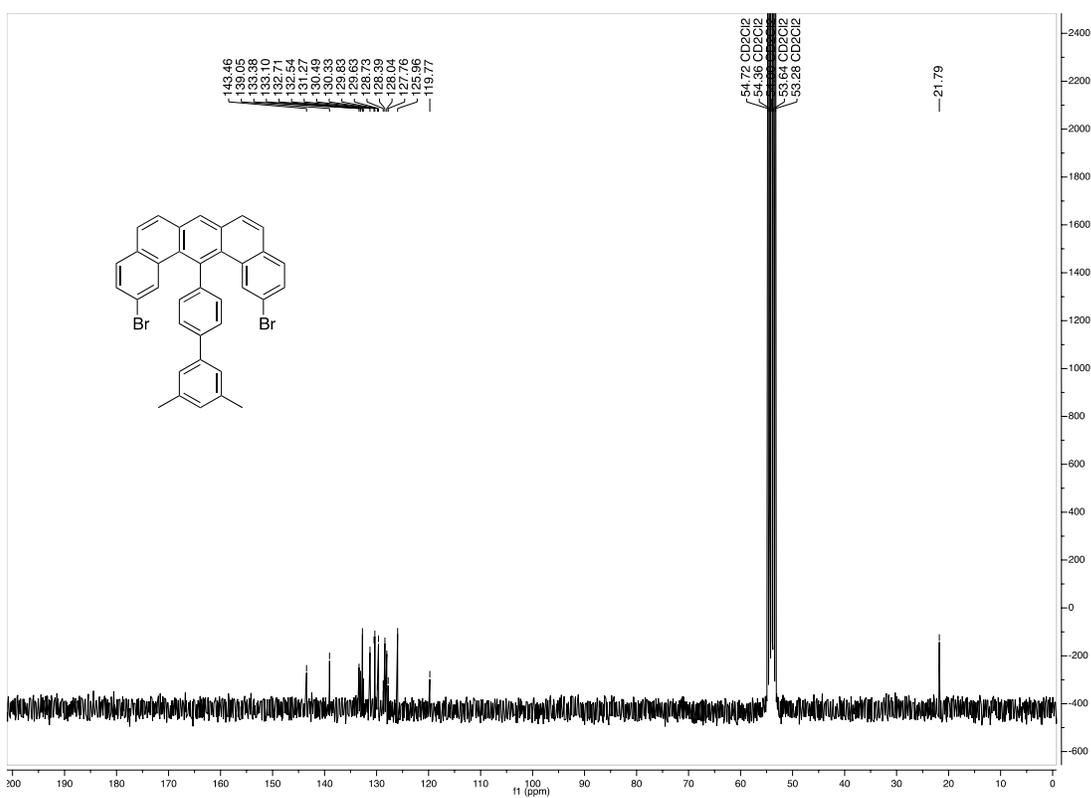

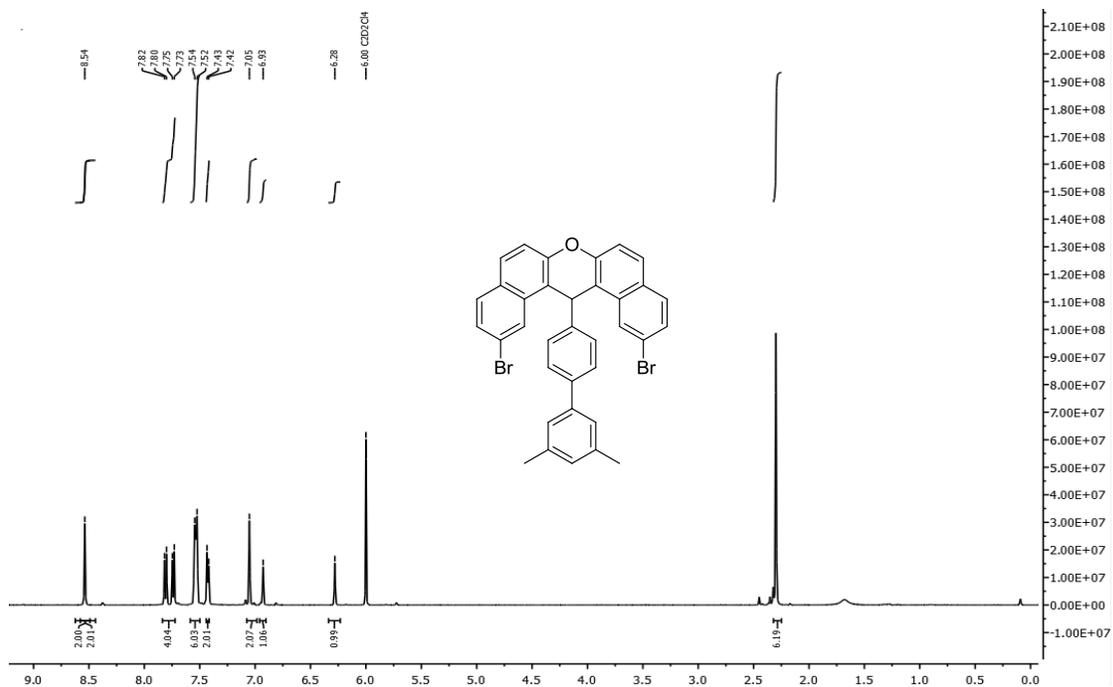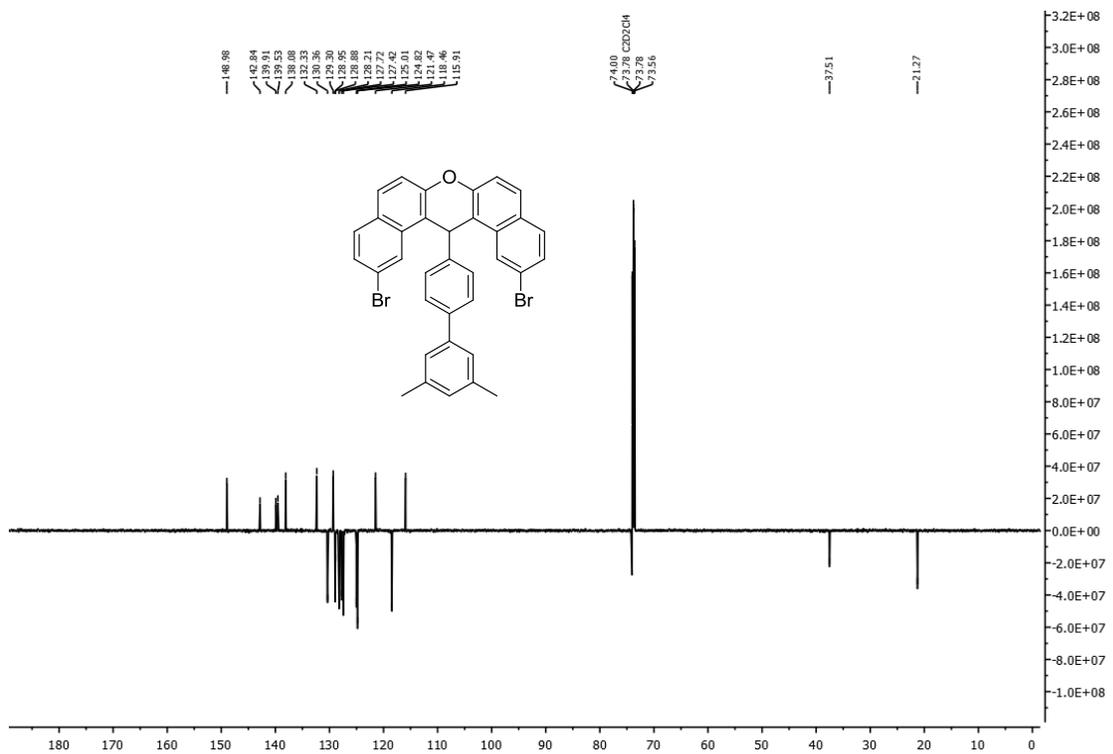

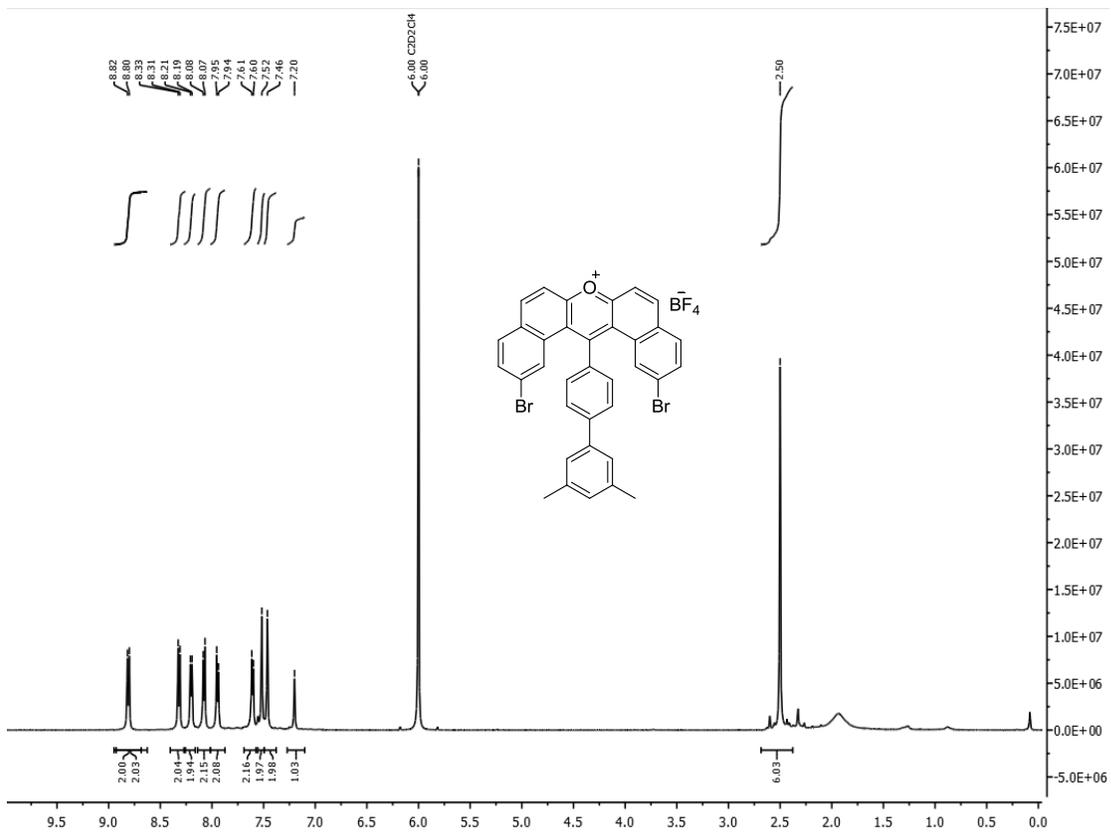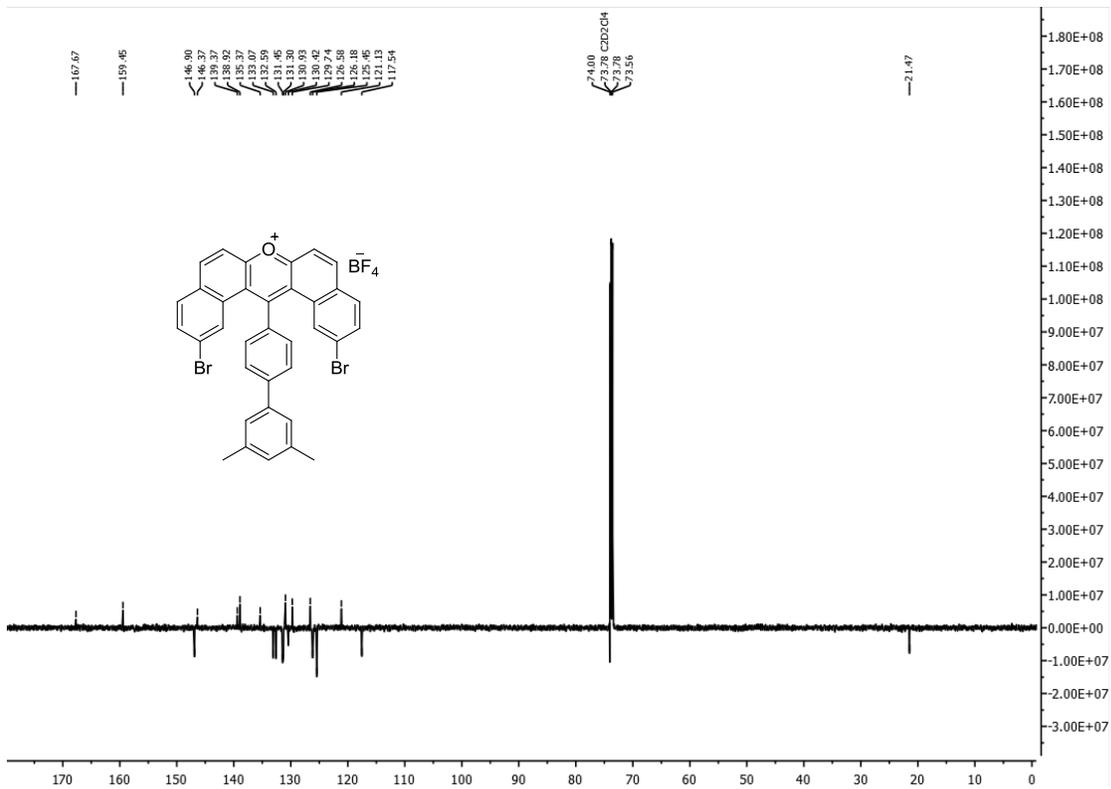

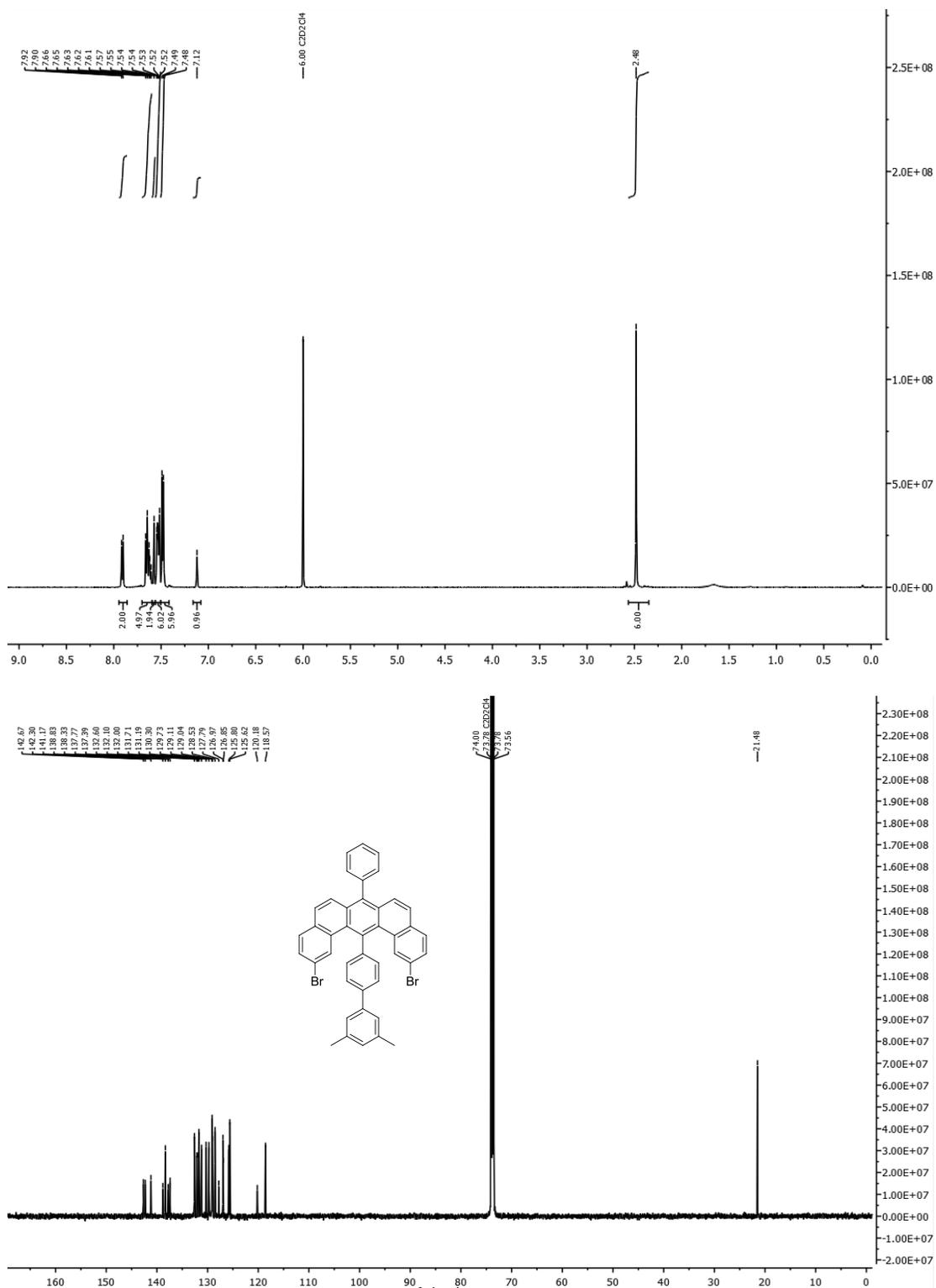

**Fig. S4.**
$^1$H and $^{13}$C NMR spectra. NMR spectra of all the intermediate products and the molecular precursors.